\documentclass[%
10pt,
twocolumn,
reprint,
nofootinbib,
nobibnotes,
amsmath,amssymb,
aps,
pra,
longbibliography
]{revtex4-1}

\usepackage{amsmath}    
\usepackage{mathtools}
\usepackage{bm}         
\usepackage{graphicx}   
\usepackage{color}      
\def\d{\mathrm{d}}

\def\b#1{\mathbf{#1}}

\def\Im{\mathrm{Im}}
\def\Re{\mathrm{Re}}
\usepackage{amssymb}
\newcommand{\vc}[1]{\ensuremath{\begin{bmatrix}#1\end{bmatrix}}}

\newcommand{\cO}{\mathcal{O}}

\begin{document}

\title{Approximate $T$-matrix and optical properties of spheroidal particles to third order in size parameter}
\author{Matt R. A. Majic}
\author{Luke Pratley}
\author{Dmitri Schebarchov}
\author{Walter R. C. Somerville}
\author{Baptiste Augui{\'e}}
\author{Eric C. Le Ru}\email{eric.leru@vuw.ac.nz}

\affiliation{The MacDiarmid Institute for Advanced Materials and Nanotechnology,
School of Chemical and Physical Sciences, Victoria University of Wellington,
PO Box 600 Wellington, New Zealand}

\date{\today}

\begin{abstract}
In electromagnetic scattering, the so-called $T$-matrix encompasses the optical response of a scatterer for any incident excitation
and is most commonly defined using the basis of multipolar fields. It can therefore be viewed as a generalization of the concept
of polarizability of the scatterer.
We here calculate the series expansion of the $T$-matrix for a spheroidal particle in the small-size/long-wavelength limit, up to third lowest order with respect to the size parameter, $\tilde X$, which we will define rigorously for a non-spherical particle.
$\mathbf{T}$ is calculated from the standard extended boundary condition method with a linear system involving two infinite matrices $\mathbf{P}$ and $\mathbf{Q}$, whose matrix elements are integrals on the particle surface. We show that the limiting form of the $P$- and $Q$-matrices, which is different in the special case of spheroid, ensures that this Taylor expansion can
be obtained by considering only multipoles of order 3 or less (i.e. dipoles, quadrupoles, and octupoles).
This allows us to obtain self-contained expressions for the Taylor expansion of $\mathbf{T}(\tilde X)$.
The lowest order is $\mathcal{O}(\tilde X^3)$ and equivalent to the quasi-static limit or Rayleigh approximation. Expressions to order $\mathcal{O}(\tilde X^5)$ are obtained by Taylor expansion of the integrals in $\mathbf{P}$ and $\mathbf{Q}$ followed by matrix inversion. We then apply a radiative correction scheme, which makes the resulting expressions valid up to order $\mathcal{O}(\tilde X^6)$.
Orientation-averaged extinction, scattering, and absorption cross-sections are then derived.
All results are compared to the exact $T$-matrix predictions to confirm the validity of our expressions and assess
their range of applicability. For a wavelength of 400\,nm, the new approximation remains valid (within 1\% error) up to
particle dimensions of the order of $100-200$\,nm depending on the exact parameters (aspect ratio and material).
These results provide a relatively simple and computationally-friendly alternative to the standard $T$-matrix method for
spheroidal particles smaller than the wavelength, in a size range much larger than for the commonly-used Rayleigh approximation.
\end{abstract}

\maketitle

\section{Introduction}

The $T$-matrix method for electromagnetic scattering can be viewed as an extension of Mie theory to non-spherical particles \cite{1965Waterman,1971WatermanPRD,1975BarberAO,2002Mishchenko}.
It provides a framework in which the optical properties of particles can be computed using a basis of spherical harmonics, whereby incident, internal, and scattered fields are expanded as a series of multipolar fields. This approach is particularly suited to predictions of orientation-averaged properties.
The $T$-matrix method has been used extensively in various contexts \cite{2004Database,2017Database},
in particular to predict the optical properties of nanoparticles beyond the (long-wavelength) Rayleigh/Gans approximation \cite{Gans}.
The latter is typically valid for particle sizes smaller than $\lambda/20-\lambda/10$, which severely limits its applicability to particles
relevant to experiments. The Rayleigh approximation has nevertheless often been used outside this range of validity \cite{1982WokaunPRL,2003KellyJPCB} for semi-quantitative calculations, simply because of the dramatic jump in complexity and computational requirements to implement the $T$-matrix framework, which provides an exact solution.
We here propose an appealing compromise between accuracy and complexity: an analytic formula for the higher-order approximation of the rigorous $T$-matrix solution, which expands the range of applicability of the approximation to a more relevant range of sizes (up to $\approx \lambda /6-\lambda /3$) whilst to some extent retaining its simplicity. Similar higher-order long-wavelength approximations have previously been obtained for spheres \cite{1982WokaunPRL,2009MorozJOSAB,2009book,2016ColomPRB,2017ColomPRA} and for spherical nano-shells  \cite{2013SchebarchovPCCP} starting from the exact solution of Mie theory. Semi-empirical approximations have also been proposed for non-spherical nanoparticles~\cite{2003KuwataAPL,2017YuCSR}, but numerical calculations are then required to pre-determine some of the parameters. Taylor expansions of the solution can also be formally
obtained by iterative solution of Laplace equation \cite{1953StevensonJAP1}. While this approach is applicable to a general shape, it results in relatively
lengthy expressions for spheroidal particles \cite{1953StevensonJAP2}, which are not easily linked to optical properties \cite{1974Heller,1977Dassios}
and do not lend themselves to orientation-averaging.
Ref.~\cite{2009MorozJOSAB} has also proposed a set of higher-order approximations for spheroids following the physical arguments
of Ref.~\cite{1983MeierOL}. We will see that these do not contain all the higher-order corrections found here.

We first review the basic ingredients of the $T$-matrix/extended boundary condition method (EBCM) for axisymmetric particles and the associated symmetry properties.
Simplified integral expressions for the auxiliary $Q$- and $P$-matrices are used to determine the dependence on the size-parameter in the small-particle limit, for all matrix elements. We show that for a general axisymmetric particle, the $T$-matrix at lowest order in the size parameter $\tilde X$ (which is $\cO(\tilde X^3)$) depends on the matrix elements in $\b P$ and $\b Q$ for {\it all} multipoles. This precludes any simple analytic approximation of $\b T$ to lowest order using the EBCM, as an infinite number of terms would have to be included.
However, as shown previously \cite{JQSRT2012,JQSRT2013}, the $Q$-matrix for {\it spheroidal} particles has a different limiting form for small $\tilde X$ and we show that in this special case, the low-order approximation of the $T$-matrix only involves the lowest multipoles, namely dipoles, quadrupoles, and octupoles for the accuracy we require here. Using this special property, we can derive a $T$-matrix approximation for a spheroidal particle up to the next non-zero correction, which is $\cO(\tilde X^5)$. To this order, only 11 independent matrix elements are non-zero. We then show that the approximations can be further improved by applying a radiative correction scheme previously described in Ref.~\cite{2013LeRuPRA},
which results in a Taylor expansion of $\b T$ correct to $\mathcal{O}(\tilde X^6)$.
The resulting expressions are relatively simple and involve no special functions or series. They provide much more accurate results than the commonly-used Rayleigh approximation for nano-spheroids.

\section{The $T$-matrix formalism}

\subsection{Definition of the $T$-matrix}

We provide a summary of the main results using notations very similar to those in Refs.~\cite{2002Mishchenko} and ~\cite{2013LeRuPRA}.
We consider the standard problem of electromagnetic scattering by a bounded scatterer, e.g. a particle.
The solutions for the incident, scattered, and internal fields are each expanded on the appropriate basis
of vector spherical wavefunctions (VSWFs), $\b M^{(i)}$ and $\b N^{(i)}$:
\begin{align} 
\b E_\mathrm{inc} & = & E_0\sum_{n,m} \left( a_{nm} \b M^{(1)}_{nm}(k_1\b r) + b_{nm}\b N^{(1)}_{nm}(k_1\b r) \right), \nonumber\\
\b E_\mathrm{sca} & = & E_0\sum_{n,m} \left( p_{nm} \b M^{(3)}_{nm}(k_1\b r) + q_{nm}\b N^{(3)}_{nm}(k_1\b r) \right), \label{eqn:vsh-expansion}\\
\b E_\mathrm{int} & = & E_0\sum_{n,m} \left( c_{nm} \b M^{(1)}_{nm}(k_2\b r) + d_{nm}\b N^{(1)}_{nm}(k_2\b r) \right), \nonumber
\end{align} 
where $k_1,k_2$ are the wavenumbers in the medium and particle respectively.
The relative refractive index $s=k_2/k_1$ may be complex for absorbing or conducting materials.
The total ($n$) and projected ($m$) angular momentum indices satisfy $|m| \leq n$, and $a_{nm}$, $b_{nm}$, $p_{nm}$, $q_{nm}$,  $c_{nm}$ and $d_{nm}$ are the expansion coefficients. Explicit definitions of $\b M^{(i)}$ and $\b N^{(i)}$, which correspond to multipolar fields, can be found in \cite{2002Mishchenko,2013LeRuPRA}, and here we just note that the superscript $(i)$ specifies which spherical Bessel function is used in the definition. Regular VSWFs that are finite at the origin are given by $i=1$ and are denoted $\mathrm{Rg}\mathbf{M}$ and $\mathrm{Rg}\mathbf{N}$ in Ref.~\cite{2002Mishchenko}, whereas the irregular VSWFs with $i=3$ are outgoing spherical waves and decay to zero in the far-field. It is worth noting that a VSWF expansion may be problematic near the surface of the scatterer but this does not affect the validity of the $T$-matrix method itself. For more information on near-fields and the so-called Rayleigh hypothesis, refer for example to Ref.~\cite{2016AuguieJO} and references therein.

Regardless of  the shape of the scatterer, the linearity of Maxwell's equations implies that the scattering coefficients $p_{nm}$ and $q_{nm}$ can each be expressed as a linear combination of the incident coefficients $a_{nm}$ and $b_{nm}$. This linear relationship can be expressed as a (block) matrix equation:
\begin{equation} \label{eqn:tmatrix}
\begin{bmatrix}
{\bf p} \\ {\bf q}
\end{bmatrix}
= \b T 
\begin{bmatrix}
{\bf a} \\ {\bf b}
\end{bmatrix}
=
\begin{bmatrix}
\b T^{11} & \b T^{12} \\ \b T^{21} & \b T^{22}
\end{bmatrix}
\begin{bmatrix}
{\bf a} \\ {\bf b}
\end{bmatrix},
\end{equation}
where the column vectors ${\bf p}$, ${\bf q}$, ${\bf a}$ and ${\bf b}$ respectively contain all the $p_{nm}$, $q_{nm}$, $a_{nm}$ and $b_{nm}$ as components; and $\b T$ is the so-called \emph{transition} matrix, or simply the $T$-matrix. Presuming that all the incident field coefficients are known, then the knowledge of the $T$-matrix will fully determine the scattered field coefficients through Eq.~\ref{eqn:tmatrix}, and therefore any derived physical properties. 

Symmetries in the particle shape can reduce some of the $T$-matrix  components to zero; if these particular components are identified \emph{a priori}, then the number of 
components that need to be computed will be reduced. This is best exemplified by spherical scatterers (which fall in the realm of Mie theory \cite{1983Bohren}): the corresponding $T$-matrix is diagonal, yielding a set of completely decoupled equations of the form $p_{nm} = \Gamma_{n}a_{nm}$ and $q_{nm} = \Delta_{n}b_{nm}$ where $\Gamma_{n}$ and $\Delta_{n}$ are the Mie coefficients \cite{2002Mishchenko}. 

\subsection{Extended Boundary Condition Method (EBCM)}

It is possible to express the elements of the $T$-matrix in semi-analytic form using the so-called Extended Boundary Condition Method (EBCM). The derivation can be found in Ref.~\cite{2002Mishchenko}. One first defines the $P$- and $Q$-matrices, which similarly to $\mathbf{T}$, express linear relationships
between the field expansion coefficients:
\begin{equation} \label{eqn:pq}
\begin{bmatrix}
{\bf a} \\ {\bf b}
\end{bmatrix}
=
\begin{bmatrix}
\b Q^{11} & \b Q^{12} \\ \b Q^{21} & \b Q^{22}
\end{bmatrix}
\begin{bmatrix}
{\bf c} \\ {\bf d}
\end{bmatrix}
\quad \textrm{and} \quad
\begin{bmatrix}
{\bf p} \\ {\bf q}
\end{bmatrix}
=
-\begin{bmatrix}
\b P^{11} & \b P^{12} \\ \b P^{21} & \b P^{22}
\end{bmatrix}
\begin{bmatrix}
{\bf c} \\ {\bf d}
\end{bmatrix}.
\end{equation}
Matrices $\b Q$ and $\b P$ relate the incident field ($\b E_\mathrm{inc}$) and the scattered field ($\b E_\mathrm{sca}$) coefficients to the internal field ($\b E_\mathrm{in}$) coefficients. The key insight of the EBCM is that the elements of $\b Q$ and $\b P$ can be expressed analytically as surface integrals of products of VSWFs, with the integration being performed over the surface of the scatterer. General expressions are given in Ref.~\cite{2002Mishchenko}.

Comparing Eqs.~\ref{eqn:tmatrix} and \ref{eqn:pq} then leads to the fundamental relationship of the EBCM method:
\begin{equation} \label{eqn:ebcm}
\b T = -\b P \b Q^{-1},
\end{equation}
which allows one to calculate the $T$-matrix elements. Note that Mishchenko \emph{et al.} write $\b P = \textrm{Rg}\b Q$, explicitly indicating that $\b P$ and $\b Q$ are mathematically very similar: they differ only in the type of spherical Bessel function used in the integral expressions, which can be traced back to the initial series expansions in Eq.~\ref{eqn:vsh-expansion}. It will be convenient for us to also define the $R$-matrix where $\b R=\b Q^{-1}$, which gives the internal series coefficients in terms of the incident coefficients. 

The $T$-, $R$-, $P$- and $Q$-matrices are infinite but in practice must be truncated at some maximum multipolarity $N$ for computations (note that we will use the term
multipolarity instead of multipole order to avoid confusion with the order of terms in our Taylor expansions).
This truncation may introduce an additional error in the $T$-matrix elements themselves, arising from the matrix inversion and multiplication in Eq. \eqref{eqn:ebcm}
\cite{JQSRT2015}.

\subsection{Axisymmetric particles with mirror symmetry}
\label{SecMirror}

For particles with symmetry of revolution around the $z$-axis, such as spheroids, expansion coefficients
with different $m$ values are entirely decoupled, and one can therefore solve the problem for each value of $m$, where $m$ can be viewed as a fixed parameter (added as a subscript where needed, but otherwise implicit).
Moreover, we have \cite{2002Mishchenko}:
\begin{align}
T^{11}_{nk|-m} &= \phantom{-}T^{11}_{nk|m}& T^{12}_{nk|-m} &= -T^{21}_{nk|m}  \nonumber\\[0.2cm]
T^{21}_{nk|-m} &= -T^{12}_{nk|m}& T^{22}_{nk|-m} &= \phantom{-}T^{22}_{nk|m} 
\end{align}
and therefore only $m\ge 0$ need to be considered in the calculation of $\mathbf{T}$.
Furthermore, the surface integrals for the $P$- and $Q$- matrix elements reduce to line integrals,
for which a number of simplified expressions have been derived \cite{2011SomervilleOL,JQSRT2012}.
The integral expressions used in this work are summarized in App.~\ref{AppIntegrals}.

Reflection symmetry with respect to the equatorial plane also results in a number of additional simplifications (see Sec.~5.2.2 of Ref.~\cite{2002Mishchenko} and Sec.~2.3 of Ref.~\cite{JQSRT2013}):
\begin{alignat}{3}
  P^{11}_{nk} &= P^{22}_{nk} &= Q^{11}_{nk} &= Q^{22}_{nk} &= 0 & \quad\text{if~}n+k\text{~odd}\nonumber, \\
  P^{12}_{nk} &= P^{21}_{nk} &= Q^{12}_{nk} &= Q^{21}_{nk} &= 0 & \quad\text{if~}n+k\text{~even},
\end{alignat}
and identical relations for $\mathbf{T}$ and $\mathbf{R}$. 
As discussed in \cite{JQSRT2013}, we can then rewrite Eqs.~\ref{eqn:tmatrix} and \ref{eqn:pq} as two independent sets of equations.
Following Ref.~\cite{2016SMARTIES}, we define
\begin{align}
\mathbf{a}_e=  \begin{pmatrix}
a_2\\a_4\\\vdots
\end{pmatrix},
\mathbf{b}_o=  \begin{pmatrix}
b_1\\b_3\\\vdots
\end{pmatrix},
\mathbf{a}_o=  \begin{pmatrix}
a_1\\a_3\\\vdots
\end{pmatrix},
\mathbf{b}_e=  \begin{pmatrix}
b_2\\b_4\\\vdots
\end{pmatrix},
\end{align}
and similarly for $\mathbf{c}$, $\mathbf{d}$, $\mathbf{p}$, $\mathbf{q}$. We also define the matrices $\mathbf{Q}^{eo}$ and $\mathbf{Q}^{oe}$
from $\mathbf{Q}$ as:
\begin{align}
\mathbf{Q}^{eo}=\begin{pmatrix}
\mathbf{Q}^{11}_{ee} & \mathbf{Q}^{12}_{eo} \\[0.4cm]
\mathbf{Q}^{21}_{oe} & \mathbf{Q}^{22}_{oo}
\end{pmatrix},\qquad
\mathbf{Q}^{oe}=\begin{pmatrix}
\mathbf{Q}^{11}_{oo} & \mathbf{Q}^{12}_{oe} \\[0.4cm]
\mathbf{Q}^{21}_{eo} & \mathbf{Q}^{22}_{ee}
\end{pmatrix},
\label{EqnBlockEOOE}
\end{align}
where $Q^{12}_{eo}$ denotes the submatrix of $Q^{12}$ with even row indices
and odd column indices, and similarly for the others. One can see that $\mathbf{Q}^{eo}$ and $\mathbf{Q}^{oe}$ contain all the non-zero elements of $\mathbf{Q}$ and exclude all the elements that must be zero by reflection symmetry, so this is an equivalent description of the $Q$-matrix.
Physically, the $eo$ matrices relate the properties of the even magnetic and odd electric multipoles, while the $oe$ matrices relate the odd magnetic and even electric multipoles. 
Those two groups are strictly decoupled in the case of axisymmetric particles with mirror symmetry and the equations relating the expansion coefficients can be written as two independent sets, for example:
\begin{equation}
\begin{pmatrix} \mathbf{a}_e\\ \mathbf{b}_o\end{pmatrix}=\mathbf{Q}^{eo}
\begin{pmatrix} \mathbf{c}_e\\ \mathbf{d}_o\end{pmatrix},\qquad
\begin{pmatrix} \mathbf{a}_o\\ \mathbf{b}_e\end{pmatrix}=\mathbf{Q}^{oe}
\begin{pmatrix} \mathbf{c}_o\\ \mathbf{d}_e\end{pmatrix},
\end{equation}
and similar expressions deduced for $\mathbf{P}$, $\mathbf{T}$, and $\mathbf{R}$. As a result, the problem of finding a $2N\times 2N$ $T$-matrix reduces to finding two decoupled $T$-matrices $\mathbf{T}^{eo}$ and $\mathbf{T}^{oe}$, each of size $N\times N$, namely:
\begin{align}
\mathbf{T}^{eo} = -\mathbf{P}^{eo}\left[\mathbf{Q}^{eo}\right]^{-1},\qquad
\mathbf{T}^{oe} = -\mathbf{P}^{oe}\left[\mathbf{Q}^{oe}\right]^{-1}.
\end{align}


\section{Size dependence of the matrix elements}

\subsection{Size parameter}
\label{SecSizeParameter}

Here we are interested in scattering by axisymmetric particles that are small relative to the wavelength. If the surface of the scatterer in spherical coordinates is described by $r(\theta)$, then we define the variables
\begin{align}
x&=k_1 r(\theta) \\
x_{\theta} & =   \frac{\d x}{\d\theta}  =k_1 \frac{\d r}{\d\theta}  \\
X&=k_1r(\theta=0) \\
\tilde X&=k_1 r_\mathrm{eq} = k_1\sqrt[3]{\frac{3V}{4\pi}}\label{eqn:xfactor}
\end{align}
where $k_{1}$ is the wavenumber in the external medium: $k_{1} = 2\pi n_1/\lambda$, with $n_{1}$ the refractive index.
$x(\theta)$ is analogous to the size parameter in Mie theory \cite{1983Bohren}, but is a function of $\theta$ for non-spherical particles.
$\tilde X$ is our chosen definition for the size parameter of a non-spherical particle, in terms of $r_\mathrm{eq}$, the radius of the sphere of equivalent volume $V$. This choice will be motivated by the fact that the range of validity of the small-size approximation depends primarily on $\tilde X$, rather than $X$.
$X$ will nevertheless be a more convenient parameter for the Taylor expansions.
For a given shape, scaling $\tilde X$ or $X$ is equivalent to scaling the particle size.

For a spheroid with symmetry of revolution around $z$, with
semi-axes $c$ along its axis and $a$ perpendicular to it,
we have:
\begin{align}
r(\theta) = {}& \frac{ac}{\sqrt{a^2\cos^2\theta + c^2\sin^2\theta}}~~\mathrm{and} ~~X=k_1 c.
\end{align}
The shape may then be characterized by a single parameter, for example the aspect ratio defined as $h=c/a$:
\begin{align}
  x(\theta) = {}& \frac{X}{\sqrt{1 + (h^2-1)\sin^2\theta}}.
\end{align}
The size is characterized by $X$ or $\tilde X$, which are related by $X = \tilde X h^{2/3}$.
For a prolate spheroid, we have $c>a$ and $h>1$, and the opposite for an oblate spheroid.
To simplify our expressions, we will also use the eccentricity of the spheroid $e=\sqrt{h^2-1}/h$.
We will carry out all our derivations for prolate spheroids (for which $0<e<1$), and discuss the extension to oblate spheroids (for which $e$ is imaginary within
the definition above) at the end.

\subsection{Long-wavelength limit of the matrix elements}

The simplified integrals (given in App.~\ref{AppIntegrals}) contain products of the Riccati-Bessel functions, for which we can develop a long-wavelength approximation by considering a Taylor series in their argument, either $x(\theta)$ or $sx(\theta)$. For example
\begin{align}
Q^{12}_{nk}&=A_nA_k\frac{s^2-1}{s} m\int^\pi_0 \d\theta  d_{n} d_{k} \xi_n(x)\psi^\prime_k(sx)x_\theta,
\end{align}
where $A_n$ is a constant, $d_n$ is a function of $\theta$ (related to the associated Legendre functions) and $\xi_n, \psi_n$ are Riccati-Bessel functions.
Since $\xi_n(x)\propto x^{-n}$ and $\psi'_k(sx)\propto (sx)^{k}$, we have $Q^{12}_{nk}\propto X^{k-n+1}$. 
Following similar arguments, the asymptotic forms of the matrix elements of the $P$- and $Q$-matrices were derived for
a general axisymmetric scatterer (see App.~\ref{AppIntOrder} and also Ref.~\cite{JQSRT2012}): 
\begin{align*}
Q^{11}_{nk} & = \cO \left( X^{k-n+2-2\delta_{nk}}\right) 
&&Q^{12}_{nk} = \cO \left( X^{k-n+1}\right),  \\   
Q^{21}_{nk} &= \cO \left( X^{k-n+1}\right),  \quad
&&Q^{22}_{nk} = \cO \left( X^{k-n}\right),  \\ \\
P^{11}_{nk} &= \cO \left( X^{k+n+3}\right),  \quad
&&P^{12}_{nk} = \cO \left( X^{k+n+2}\right), \\
P^{21}_{nk} &= \cO \left( X^{k+n+2}\right),   \quad
&&P^{22}_{nk} = \cO \left( X^{k+n+1}\right).
\end{align*}

From these, it is possible to prove (see App.~\ref{AppInvGen} for details)
that ${\b R} ={\b Q}^{-1}$ and ${\b T} = -{\b P}{\b Q}^{-1}$ have the same lowest order dependence as $\b Q$ and $\b P$ respectively:
\begin{align}
R^{ij}_{nk} \propto Q^{ij}_{nk} 
~~\mathrm{and}~~ T^{ij}_{nk} \propto P^{ij}_{nk} 
\quad\mathrm{For~} i,j =1,2.
\end{align}
Following this proof, one may notice that to obtain the lowest order approximation of
$\b R$ or $\b T$ (in particular $T^{22}_{11}$, akin to the electric dipole polarizability of the particle), it is necessary to include elements of of $\b P$ and $\b Q$ of higher multipolarity.
This important point is more obvious when writing out explicitly the lowest-order form of the matrices. For example, for a mirror-symmetric particle, we have in the small particle limit (truncated at $N=5$):
\begin{align}
\mathbf{P}^{eo}, \mathbf{T}^{eo}&=\cO 
\left[\begin{array}{cc|ccc}
X^{ 7} & X^{ 9} & X^{ 5} & X^{ 7} & X^{ 9 }\\
X^{ 9} & X^{ 11} & X^{ 7} & X^{ 9} & X^{ 11} \\
\hline
\color{blue} X^{ 5} & \color{blue} X^{ 7} & \color{blue} 
X^{ 3} & \color{blue} X^{ 5} & \color{blue} X^{ 7} \\
X^{ 7} & X^{ 9} & X^{ 5} & X^{ 7} & X^{ 9} \\
X^{ 9} & X^{ 11} & X^{7} & X^{ 9} & X^{11}
\end{array}\right],
 \label{EqnTeo}
\\[0.5cm]
\mathbf{Q}^{eo}, \mathbf{R}^{eo}&=\cO 
\left[\begin{array}{cc|ccc}
X^{ 0} & X^{ 4} & \color{blue} X^{ 0} & X^{ 2} & X^{ 4 }\\
X^{ 0} & X^{ 0} & \color{blue} X^{-2} & X^{ 0} & X^{ 2} \\
\hline
X^{ 2} & X^{ 4} & \color{blue} X^{ 0} & X^{ 2} & X^{ 4} \\
X^{ 0} & X^{ 2} & \color{blue} X^{-2} & X^{ 0} & X^{ 2} \\
X^{-2} & X^{ 0} & \color{blue} X^{-4} & X^{-2} & X^{ 0}
\end{array}\right].
\end{align}
We see that although the dominant term $T^{22}_{11}$ is of order $\cO(X^3)$, it will be necessary to compute the lowest order terms for the entire column of ${\b R}^{eo}=[\b Q^{eo}]^{-1}$
and the entire row of $\b P^{eo}$ in order to compute it from $\b T^{eo}=-\b P^{eo}\b R^{eo}$.

\subsection{The curious case of spheroidal particles}
\label{SecSpheroids}

As shown in Ref.~\cite{JQSRT2012}, the matrix elements of the lower triangular parts of the four blocks of the $Q$-matrix have a different limiting form in the special case of spheroidal particles. All terms of negative orders in $X$ in the Taylor expansion of the $Q$-matrix are exactly zero, resulting in $Q^{ij}_{nk}=\cO(X^0)$ for $n>k$ and $i,j=1,2$. As a result, it can be proved that the same applies to $\b R$, i.e. $R^{ij}_{nk}=\cO(X^0)$ for $n>k$ (see App.~\ref{AppInvSph}). 
Explicitly for the ${eo}$ matrices, we then have:
\begin{equation}
\mathbf{Q}^{eo}, \mathbf{R}^{eo}=\cO 
\left[\begin{array}{cc|ccc}
X^{ 0} & X^{ 4} & X^{ 0} & X^{ 2} & X^{ 4 }\\
X^{ 0} & X^{ 0} & X^{ 0} & X^{ 0} & X^{ 2} \\
\hline
X^{ 2} & X^{ 4} & \color{blue} X^{ 0} & X^{ 2} & X^{ 4} \\
X^{ 0} & X^{ 2} & X^{ 0} & X^{ 0} & X^{ 2} \\
X^{ 0} & X^{ 0} & X^{ 0} & X^{ 0} & X^{ 0}
\end{array}\right].
\end{equation}
Together with Eq.~\ref{EqnTeo}, this implies that only $R^{22}_{11}$ will contribute to the lowest order of $T^{22}_{11}$.
Similarly, when calculating $\b T^{eo}= - \b P^{eo} \b R^{eo}$ to order $\cO(X^5)$, we may then truncate all three matrices to multipolarity $N=3$, i.e. consider only the electric dipole and octupole and the magnetic quadrupole. All matrix elements in $\b P^{eo}$ and $\b R^{eo}$ can moreover be approximated by their lowest order term
except $P^{22}_{11}$ and $R^{22}_{11}$, which must be expanded up to their next non-zero term ($X^5$ and $X^2$, respectively).
Moreover, by tracking the order of terms in the inversion of $\b Q$, one can also show that $\b Q$ can also be truncated at $N=3$ when
calculating $\b T$ to order $\cO(X^5)$, see App.~\ref{AppInvTrunc} for details.
In other words, all matrix elements with $n \geq 4$ or $k \geq 4$ will ultimately only introduce corrections of order $\cO(X^7)$ or higher for $\b T$.

The situation is even simpler in the $oe$ case as the lowest order elements of $\b P^{oe}$ and $\b T^{oe}$
are $\cO(X^5)$:
\begin{align}
\mathbf{P}^{oe}, \mathbf{T}^{oe}&=\cO 
\left[\begin{array}{ccc|cc}
X^{ 5} & X^{ 7} & X^{ 9} & X^{ 5} & X^{ 7 }\\
X^{ 7} & X^{ 9} & X^{11} & X^{ 7} & X^{ 9} \\
X^{ 9} & X^{11} & X^{13} & X^{ 9} & X^{11} \\
\hline
X^{ 5} & X^{ 7} & X^{ 9} & X^{ 5} & X^{ 7} \\
X^{ 7} & X^{ 9} & X^{ 11} & X^{ 7} & X^{9}
\end{array}\right].
\end{align}

The same cancellations occur in the lower triangular part of $Q^{oe}$ in the case of spheroids, giving:
\begin{align}
\mathbf{Q}^{oe}, \mathbf{R}^{oe}&=\cO 
\left[\begin{array}{ccc|cc}
X^{ 0} & X^{ 4} & X^{ 6} & X^{ 2} & X^{ 4 }\\
X^{ 0} & X^{ 0} & X^{ 4} & X^{ 0} & X^{ 2} \\
X^{ 0} & X^{ 0} & X^{ 0} & X^{ 0} & X^{ 0} \\
\hline
X^{ 0} & X^{ 2} & X^{4} & X^{ 0} & X^{ 2} \\
X^{ 0} & X^{ 0} & X^{2} & X^{ 0} & X^{ 0}
\end{array}\right].
\end{align}
Following similar arguments as for the $eo$ matrices, we can show that in order to obtain $\b T^{oe}$ to order $\cO(X^5)$,
we can truncate all $oe$ matrices at multipolarity $N=2$.

Overall, we see that we can ignore any matrix elements related to electric multipolarity larger than 3 or to magnetic multipolarity larger than 2.

\subsection{Radiative correction}

The final ingredient of our approach will be to enforce a suitable radiative correction following Ref.~\cite{2013LeRuPRA}.
Briefly, this procedure consists in defining the matrix $\mathbf{U}$ such that $\b Q = \b P + i\b U$.
$\b U$ is computed exactly like $\b Q$, but replacing the Riccati-Hankel functions $\xi_n = \psi_n + i\chi_n$
by the irregular Riccati-Bessel functions $\chi_n$.
We can then compute the so-called $K$-matrix defined as $\b K = \b P \b U^{-1}$ \cite{2013LeRuPRA}, which therefore has a similar expression as the $T$-matrix in terms of $\b Q$. $i\chi_n$ and $\xi_n$ have the same lowest-order expansion and therefore $\b K$ has the same long-wavelength behaviour as $\b T$.
The final step in this scheme is to compute $\b T$ from:
\begin{align}
\b T = - \left[\b I + i \b K^{-1}\right]^{-1} = i\b K (\b I -i \b K)^{-1}
\label{EqnRC}
\end{align}

This procedure therefore requires an additional matrix inversion step, but the resulting expressions are more physically valid especially in the case
of non-absorbing particles \cite{2013LeRuPRA}.
An additional benefit here is that the Taylor expansion of $\chi_n$ (and therefore of $\b U$) only contains even {\it or} odd orders, while
that of $\xi_n$ contains both. As a result, more terms in the Taylor expansions of the matrix elements will reduce to zero.

\subsection{Comparison to the analytic quasi-static solution}

The lowest-order approximation to the matrix elements can alternatively be obtained by solving
the boundary value problem in the quasi-static approximation. Analytic results have recently been derived for the lower-right blocks ($\b P^{22}$, $\b Q^{22}$, $\b T^{22}$)  \cite{2017MajicJQSRT} where the problem can be solved using separation of variables in spheroidal coordinates. Similar expressions have been found for the other matrix blocks although these results have not yet been published. Where applicable, this method may be simpler than Taylor-expanding the integrals to lowest order and will provide us with equivalent, and sometimes simpler, expressions.

\section{Taylor expansion of the $T$-matrix and derived quantities to $\cO(X^6)$}

We now bring together all the preliminary results of the previous section to calculate the Taylor expansions of the $T$-matrix up to order $\cO(X^6)$
for spheroidal particles.
Firstly, as explained in Sec.~\ref{SecSpheroids}, we can truncate all matrices at multipole $N= 3$, as higher-multipole matrix
elements will only introduce corrections of order $\cO(X^7)$ or higher.
Since $|m|\le n$, we only need to consider $0\le m \le 3$ (in fact we will find that only up to $m=2$ are relevant). Since different $m$ are decoupled, it is easier to study each $m$ separately.
We will start with $m=0$ and provide full details of the derivations. We then provide the final results with fewer details for $m=1$ and 2.
For each $m$ we proceed as follows:
\begin{itemize}
\item
Formally write out the Taylor expansions of the matrix elements of $\b P$ up to order $X^5$ and $\b U$ up to order $X^2$.
\item
Deduce the Taylor expansion of $\b K = \b P \b U^{-1}$ up to order $X^5$ as a function of the expansion coefficients of $\b P$ and $\b U$.
\item
Calculate explicit expressions for the necessary $\b P$ and $\b U$ expansion coefficients by expanding the integrals defining the $P$- and $U$- matrix elements.
\item
Simplify expressions as much as possible to obtain a concise expression for $\b K$ to order $X^5$.
\item
Apply the radiative correction to deduce $\b T$ (which will then be valid to order $X^6$).
\end{itemize}
All results will be expressed using a mixture of the aspect ratio $h$ and eccentricity $e$.

\subsection{$K$-matrix for $m=0$}

For $m=0$ the off diagonal blocks $\b U^{12}, \b U^{21}$, $\b P^{12}, \b P^{21}$, $\b K^{12}, \b K^{21}$ are zero.
The ${eo}$ matrices are truncated at $N=3$ and are of the form
\begin{equation}
\mathbf{P}^{eo}_{m=0} =  \left[\begin{array}{c|cc}
P^{11}_{22} & 0 & 0\\
\hline
0 & P^{22}_{11} & P^{22}_{13} \\
0 & P^{22}_{31} & P^{22}_{33}
\end{array}\right].
\end{equation}
The matrices therefore decouple into $2\times 2$ and $1\times 1$ blocks. The $1\times 1$ block can be ignored entirely as
$K^{11}_{22} = \cO(X^7)$ so we only need to consider $\b K^{22}$, which can be obtained from
\begin{equation}
\vc{
K^{22}_{11} & K^{22}_{13} \\
K^{22}_{31} & K^{22}_{33} } =
\vc{
P^{22}_{11} & P^{22}_{13} \\
P^{22}_{31} & P^{22}_{33} }
\vc{
U^{22}_{11} & U^{22}_{13} \\
U^{22}_{31} & U^{22}_{33} }^{-1} .
\label{EqnKPU}
\end{equation}

We now expand the matrix elements as power series in $X$, making use of the fact that only even or odd orders in the Taylor expansions of $\psi_n$ and $\chi_n$ are non-zero:
\begin{align}
\vc{
P^{22}_{11} & P^{22}_{13} \\
P^{22}_{31} & P^{22}_{33} } =
X^3 \vc{
p_{11}+p_{11}^{(2)} X^2 & p_{13} X^2\\
p_{31} X^2& 0 } +\cO(X^7).
\end{align}
The coefficients $p_{nk}$ are by construction independent of $X$.


It is less obvious to see which terms should be kept in the matrix $\b U^{22}$ in order to correctly obtain the terms required
in $\b K^{22} = \b P^{22} [\b U^{22}]^{-1}$, but, noting that $\b U^{-1}$ has the same small $X$ dependence as $\b U$, one can show (see App.~\ref{AppInvU}) that we only need:
\begin{align}
\vc{
U^{22}_{11} & U^{22}_{13} \\
U^{22}_{31} & U^{22}_{33} } =
\vc{
u_{11}+u_{11}^{(2)} X^2 & u_{13} X^2 \\
u_{31} & u_{33} }
+\cO(X^4)
\end{align}
All higher order terms will only contribute to $\cO(X^7)$ corrections in $\b K^{22}$.
The $K$-matrix is obtained from Eq.~\ref{EqnKPU} as:
\begin{align}
\vc{
K^{22}_{11} & K^{22}_{13} \\
K^{22}_{31} & K^{22}_{33} } =
\vc{
K^{22}_{11|0} &  \left(\dfrac{p_{13}}{u_{33}} - \dfrac{p_{11}u_{13}}{u_{11}u_{33}}\right)  X^5\\
\dfrac{p_{31}}{u_{11}}X^5 & 0 } 
+\cO(X^7),
\label{EqnKm0}
\end{align}
where
\begin{align}
K_{11|0}^{22} = X^3  \frac{K_0}{1-\Omega_0 X^2}+\cO(X^7),\label{EqnK2211m0}
\end{align}
with\begin{align}
K_0&=\frac{p_{11}}{u_{11}}, \\
\Omega_0 &=\dfrac{p_{11}^{(2)}}{p_{11}}-\dfrac{u_{11}^{(2)}}{u_{11}}+ \dfrac{u_{31}}{u_{33}}\left(\dfrac{u_{13}}{u_{11}}- \dfrac{p_{13}}{p_{11}}\right).
\label{EqnOmega0}
\end{align}
The dominant term $K_0 X^3$ is analogous to the static dipolar polarizability and scales with volume as expected.
We chose to write the second-order correction as a factor $(1-\Omega_0 X^2)^{-1}$ instead of the equivalent $(1+\Omega_0 X^2)$. This choice simplifies the expressions later when applying the radiative correction to obtain the $T$-matrix, and also provides a slightly better approximation.

All coefficients $p_{nk}$ and $u_{nk}$ can be derived by Taylor expansions of the corresponding integrals for the matrix elements of $\b P$ and $\b U$. Some of them can also be directly obtained from the analytic expressions of the $P$- and $Q$-matrix in the quasi-static approximation \cite{2017MajicJQSRT}. This is the case for $p_{11}$, $p_{31}$, $p_{13}$, $u_{11}$, $u_{13}$, and $u_{33}$ (but not of $u_{31}$ as $U^{22}_{31}$ reduces to zero for spheroids in the quasi-static approximation). $p_{nk}$ and $u_{nk}$ are given explicitly in App.~\ref{AppCoefm0}, and we here only focus on the final simplified expressions.

We first introduce $L_z$, the static depolarization factor for a prolate spheroid \cite{1983Bohren}, to emphasize the connection with the quasi-static limit:
\begin{align}
L_z = \frac{1-e^2}{e^2} \left[\frac{\text{atanh}(e)}{e}- 1\right].
\end{align}
The dominant term can then be expressed as:
\begin{align}
K_0 = \frac{2}{9h^2} \frac{s^2-1}{1+(s^2-1)L_z} = \frac{2}{3c^3} \frac{\alpha_{zz}}{4\pi\epsilon_0}.
\end{align}
where the proportionality to the static dipolar polarizability $\alpha_{zz}$ is shown explicitly.

Moreover, the matrix $\b K^{22}$ should be symmetric when $s$ is real \cite{2013LeRuPRA}, which is not immediately obvious from \eqref{EqnKm0}, but the equality of the off-diagonal terms can be proved by substituting the expressions of the expansion coefficients. This equality can also be used to simplify the expression for $\Omega_0$ in \eqref{EqnOmega0}, 
which after some algebra can be reduced to a relatively simple expression:
\begin{align}
\Omega_0 = \frac{1}{25 }\frac{4e^2 -5+(s^2-1)\left[5(1-e^2)+9e^2L_z\right]}{1+(s^2-1)L_z}.
\end{align}
The off diagonal terms also simplify to give:
\begin{align}
K^{22}_{13|0}=K^{22}_{31|0}&=\frac{2e^2\sqrt{14}}{1575h^2}\frac{s^2-1}{1+(s^2-1)L_z} X^5+ \cO(X^7),
\end{align}
which coincide with the quasi-static limit in \cite{2017MajicJQSRT}.

The situation is simpler for the ${oe}$-matrices as they are $2\times 2$ diagonal when truncated at $N=2$.
The matrix inversion is then trivial and we simply have for the $oe$-matrix elements:
\begin{align}
K^{11}_{11|0} &= \cO(X^5) = \frac{P^{11}_{11|0}}{U^{11}_{11|0}} + \cO(X^7), \label{EqnM01}
\\[0.2cm]
K^{22}_{22|0} &= \cO(X^5) = \frac{P^{22}_{22|0}}{U^{22}_{22|0} }+ \cO(X^7).
\label{EqnM02}
\end{align}
Since they are of order $\cO(X^5)$, we only need to take the lowest-order approximation of the numerator and denominator
in these expressions.
We obtain after simplifications:
\begin{align}
K^{11}_{11|0}&=\frac{s^2-1}{45 h^4} X^5+ \cO(X^7),\\[0.2cm]
K^{22}_{22|0}&=\frac{3-e^2}{225 h^2}\frac{s^2-1}{1+(s^2-1)L_{20}}X^5 + \cO(X^7),
\end{align}
where $L_{20}$ is a generalization of the concept of depolarization factor to quadrupolar excitation
and is a purely geometric property:
\begin{align}
L_{20} 
	   &= \frac{3}{2} \frac{1-e^2}{e^3}\bigg[\frac{3-e^2}{e^2}\text{atanh}(e) -\frac{3}{e}\bigg].
\end{align}
We will later define two more of these quadrupole factors for $m=1,2$ (in a similar fashion as in Ref.~\cite{2010GuzatovLP}). The notation is $L_{nm}$ so that we would have $L_z\equiv L_{10}$ and $L_x= L_y \equiv L_{11}$.

This completes the Taylor expansion of the $K$-matrix for $m=0$. In summary:

\begin{widetext}
\begin{align}
\b K_{m=0} = 
\left[\def\arraystretch{2.2}\begin{array}{cc|ccc}
\dfrac{s^2-1}{45 h^4} X^5 & 0 & 0 & 0 & 0\\
0 & 0 & 0 & 0 & 0\\
\hline
0 & 0 & \dfrac{K_0 X^3}{1-\Omega_0 X^2} & 0 & \dfrac{e^2\sqrt{14}}{175}K_0 X^5\\
0 & 0 & 0 & \dfrac{3-e^2}{225 h^2}\dfrac{s^2-1}{1+(s^2-1)L_{20}}X^5 & 0\\
0 & 0 & \dfrac{e^2\sqrt{14}}{175}K_0 X^5 & 0 & 0
\end{array}\right]
+\cO(X^7).
\label{EqnKm0Summary}
\end{align}
\end{widetext}

\subsection{Radiative correction and $T$-matrix for $m=0$}

We can now obtain the $T$-matrix from the $K$-matrix using Eq.~\ref{EqnRC} (i.e. the radiative correction).
This will ensure that the resulting $T$-matrix, despite being an approximation, better satisfies energy conservation constraints \cite{2013LeRuPRA}.
For the elements that are completely decoupled from others (i.e. belonging to a $1\times 1$ block), this correction is straightforward:
\begin{align}
T^{11}_{11|0}&=\frac{i K^{11}_{11|0}}{1- i K^{11}_{11|0}}+ \cO(X^7), \label{EqnTm0RC1}
\\[0.2cm]
T^{22}_{22|0}&=\frac{i K^{22}_{22|0}}{1- i K^{22}_{22|0}}+ \cO(X^7). \nonumber
\end{align}

For the $2\times 2$ block of $\b K^{22}$, the exact application of the radiative correction on the approximate $K$-matrix requires the inversion
of a $2\times 2 $ matrix and gives:
\begin{align}
&T^{22}_{11|0} = \frac{iK_0 X^3 \left[1 + i \frac{14e^4}{30625} K_0 X^{7}(1-\Omega_0 X^2)\right]}{1-\Omega_0 X^2 - i K_0 X^3\left[1 + i \frac{14e^4}{30625} K_0 X^{7}(1-\Omega_0 X^2)\right]}, \\[0.5cm]
&T^{22}_{31|0} = T^{22}_{13|0} = \\
&~~\frac{i \frac{\sqrt{14}e^2}{175} K_0 X^5 (1-\Omega_0 X^2)}{1-\Omega_0 X^2 - i K_0 X^3\left[1 + i \frac{14e^4}{30625} K_0 X^{7}(1-\Omega_0 X^2)\right]}.\nonumber
\end{align}

We could simplify these expressions by removing all terms of order $\cO(X^7)$ or more.
In doing so, we need to keep in mind that for non-absorbing particles $\b K^{11}$ and $\b K^{22}$ are real,
and $\b K^{12}$ and $\b K^{21}$ are pure imaginary matrices.
The imaginary and real parts of $\b T$ then have very different dominant terms.
For $T^{22}_{11|0}$ for example, $\Im\{T^{22}_{11|0}\} \sim X^3$ but 
$\Re\{T^{22}_{11|0}\} \sim X^6$. Because the latter is related to the extinction cross-section, it is important
to keep the dominant terms in the real part as well as the imaginary part.
For other $T$-matrix elements which are $\sim X^5$, their real part is $\sim X^{10}$, as evident for example in Eq.~\ref{EqnTm0RC1}.
If we were to neglect the $\cO(X^{10})$ terms and write for example $T^{11}_{11|0}=i K^{11}_{11|0} + \cO(X^7)$, then $T^{11}_{11|0}$ would be pure imaginary and the corresponding predicted extinction cross-sections (proportional to Re$(T)$) would be zero despite the fact that the scattering cross-sections ( $\propto |T|^2$) is non zero, a clearly non-physical result.
It is therefore important to keep the dominant terms of the real part of $\b T$ (for real $s$), even if they are of higher order than our desired approximation $\cO(X^6)$.
Following this principle, we therefore have the following simplified expressions:
\begin{align}
T^{22}_{11|0} &= \frac{iK_0 X^3}{1-\Omega_0 X^2 - i K_0 X^3} +\cO(X^7),\\[0.5cm]
T^{22}_{31|0} &= T^{22}_{13|0} = \frac{i\sqrt{14}e^2}{175}\frac{ K_0 X^5}{1-i K_0 X^3} + \cO(X^7).
\end{align}
While these are not strictly energy conserving as some terms have been neglected, the condition
(extinction = scattering + absorption) will nevertheless be approximately valid within the range of validity
of our expansions, even for non-absorbing scatterers.
Those expressions will be compared to the predictions without radiative correction, for which the
$T$-matrix is then simply given as $\b T = i \b K$.
Refer to Ref.~\cite{2013LeRuPRA} for further discussion of the radiative correction.

In summary, the $T$-matrix after radiative correction is given by:
\begin{widetext}
\begin{align}
\b T_{m=0} = 
\left[
\def\arraystretch{2.2}
\begin{array}{cc|ccc}
\dfrac{i K^{11}_{11}}{1- i K^{11}_{11}}  & 0 & 0 & 0 & 0\\
0 & 0 & 0 & 0 & 0 
\\ \hline 
0 & 0 & \dfrac{iK_0 X^3}{1-\Omega_0 X^2-iK_0 X^3} & 0 & \dfrac{\sqrt{14}}{175}\dfrac{i e^2 K_0 X^5}{1-i K_0 X^3}\\
0 & 0 & 0 & \dfrac{i K^{22}_{22}}{1- i K^{22}_{22}} & 0\\
0 & 0 & \dfrac{\sqrt{14}}{175}\dfrac{i e^2 K_0 X^5}{1-i K_0 X^3} & 0 & 0
\end{array}
\right]
+\cO(X^7).
\label{EqnTm0Summary}
\end{align}
\end{widetext}

\subsection{$K$-matrix for $m=1$}

For $m=1$ the expressions are more complicated due to coupling of the matrix elements from different matrix blocks.
Details of the derivation are given in App.~\ref{AppInvU} and we here only provide the final simplified expressions.

For the dominant dipolar polarizability term, we have
\begin{align}
K_{11|1}^{22}= \frac{K_1 X^3}{1-\Omega_1 X^3} +\cO(X^7),
\end{align}
where
$K_1$ is related to the static dipolar polarizability along the $x$ or $y$ axis, $\alpha_{xx}=\alpha_{yy}$:
\begin{align}
K_{1} = \frac{2}{9h^2} \frac{s^2-1}{1+(s^2-1)L_x} = \frac{2}{3c^3} \frac{\alpha_{xx}}{4\pi\epsilon_0}.
\end{align}
$L_x$  is the standard depolarization factor (note that\\
 {$2L_x+L_z = 1$}):
\begin{align}
L_x = \frac{-1}{2e^2}\left[\frac{1-e^2}{e}\text{atanh}(e)-1\right].
\end{align}
$\Omega_1$ is obtained as a fairly complicated expression when expressed in terms of the elements of $\b P$ and $\b U$, but like $\Omega_0$, can be substantially simplified to:
\begin{align}
\Omega_1 = \frac{1}{25} \frac{3e^2 -5 +(s^2-1)(5-12e^2L_x)}{1+(s^2-1)L_x}.
\end{align}

The other matrix elements for $\b K^{eo}$ are:
\begin{align}
K^{22}_{13|1}=K^{22}_{31|1}&= \frac{2e^2\sqrt{21}}{525} K_1 X^5+\cO(X^7), \label{K22131}\\[0.2cm]
K^{21}_{12|1}=-K^{12}_{21|1}&=\frac{i e^2\sqrt{15}}{150} K_1 X^5+\cO(X^7).\label{K21121}
\end{align}

For the $\b K^{oe}$ matrix, we again define a generalized depolarization factor:
\begin{align}
L_{21} 
& = -\frac{2-e^2}{2e^4}\left[3\frac{1-e^2}{e}\text{atanh}(e) - 3+2e^2\right],
\end{align}
and we then have
\begin{align}
K^{11}_{11|1}&=  \frac{(s^2-1)\left[h^2 (2-e^2)^2 + 4(s^2-1)L_{21}\right]}{90h^4(2-e^2)\left[1+(s^2-1)L_{21}\right]} X^5\nonumber\\
&\qquad \qquad\qquad\qquad\qquad\qquad\qquad+ \cO(X^7),\label{K11111} \\[0.2cm]
K^{22}_{22|1}&=\frac{2-e^2}{150h^2}\frac{s^2-1}{1+(s^2-1)L_{21}} X^5 + \cO(X^7)\label{K22221},\\[0.2cm]
K^{21}_{21|1}&=-K^{12}_{12|1}=   \frac{i e^2 X^5}{30\sqrt{15}h^2} \,\frac{s^2-1}{1+(s^2-1)L_{21}} + \cO(X^7). \label{K21211}
\end{align}

In summary:
\begin{widetext}
\begin{align*}
&\b K_{m=1} = \\
&\left[\def\arraystretch{2.2}\begin{array}{cc|ccc}
\frac{(s^2-1)\left[h^2 (2-e^2)^2 + 4(s^2-1)L_{21}\right]}{90h^4(2-e^2)\left[1+(s^2-1)L_{21}\right]} X^5 & 0 &
0 & \dfrac{-ie^2 X^5}{30\sqrt{15}h^2}\dfrac{s^2-1}{1+(s^2-1)L_{21}}  & 0\\
0 & 0 & -\dfrac{i e^2\sqrt{15}}{150} K_1 X^5 & 0 & 0\\[0.2cm]
\hline
0 & \dfrac{i e^2\sqrt{15}}{150} K_1 X^5 &
\dfrac{K_1 X^3}{1-\Omega_1 X^2} & 0 & \dfrac{2e^2\sqrt{21}}{525}K_1 X^5\\
\dfrac{ie^2 X^5}{30\sqrt{15}h^2}\dfrac{s^2-1}{1+(s^2-1)L_{21}}   & 0 & 0 & \dfrac{2-e^2}{150h^2}\dfrac{s^2-1}{1+(s^2-1)L_{21}}X^5  & 0\\
0 & 0 & \dfrac{2e^2\sqrt{21}}{525} K_1 X^5 & 0 & 0
\end{array}\right]&& \\
& && \hspace{-2cm} +\cO(X^7).
\end{align*}
\end{widetext}

\subsection{Radiative correction and $T$-matrix for $m=1$}

The radiative correction for the $eo$ part of the matrix is very similar to the $m=0$ case and following the
same method and arguments, we obtain:
\begin{align}
T^{22}_{11|1} &= \frac{iK_1 X^3}{1-\Omega_1 X^2 - i K_1 X^3} +\cO(X^7)\\[0.5cm]
T^{21}_{12|1} &= - T^{12}_{21|1} = \frac{i K^{21}_{12|1}}{1-i K_1 X^3} + \cO(X^7).\\[0.5cm]
T^{22}_{31|1} &= T^{22}_{13|1} = \frac{i K^{22}_{31|1}}{1-i K_1 X^3} + \cO(X^7).
\end{align}

For the $oe$ matrix, this time we have four elements (i.e. a $2\times 2$ matrix), which are all $\cO(X^5)$.
Applying the matrix inversion and keeping the dominant terms for both real and imaginary parts, we obtain:
\begin{align}
T^{11}_{11|1} &= \frac{iK^{11}_{11|1}}{1- i \left[K^{11}_{11|1} - (K^{21}_{21|1})^2/K^{11}_{11|1}\right]} +\cO(X^7)\label{T11111}\\[0.5cm]
T^{22}_{22|1} &= \frac{iK^{22}_{22|1}}{1- i \left[K^{22}_{22|1} - (K^{21}_{21|1})^2/K^{22}_{22|1}\right]} +\cO(X^7)\label{T22221}\\[0.5cm]
T^{21}_{21|1} &= - T^{12}_{12|1} = \frac{i K^{21}_{21|1}}{1-i\left[K^{11}_{11|1} + K^{22}_{22|1}\right]} + \cO(X^7). \label{T21211}
\end{align}

The denominators in (\ref{T11111}-\ref{T21211}) are interesting in their own right in terms of discussing the radiative correction, as
they are different from the obvious radiative corrections of the type $iK/(1-iK)$ encountered to date.
In fact, when comparing the predictions to exact numerical results, we can confirm that the expressions above
are the correct ones as any other choices result in problems for non-absorbing particles.

In summary:
\begin{widetext}
\begin{align*}
&\b T_{m=1} = \\ 
&\left[\def\arraystretch{2.2}\begin{array}{cc|ccc}
\dfrac{iK^{11}_{11}}{1- i \left[K^{11}_{11} - (K^{21}_{21})^2/K^{11}_{11}\right]}  & 0 &
0 & -\dfrac{i K^{21}_{21}}{1-i\left[K^{11}_{11} + K^{22}_{22}\right]} & 0\\
0 & 0 & -\dfrac{i K^{21}_{12}}{1-i K_1 X^3} & 0 & 0\\[0.2cm]
\hline
0 & \dfrac{i K^{21}_{12}}{1-i K_1 X^3}  &
\dfrac{iK_1 X^3}{1-\Omega_1 X^2 - i K_1 X^3} & 0 & \dfrac{i K^{22}_{31}}{1-i K_1 X^3}\\
\dfrac{i K^{21}_{21}}{1-i\left[K^{11}_{11} + K^{22}_{22}\right]}  & 0 & 0 & \dfrac{iK^{22}_{22}}{1- i \left[K^{22}_{22} - (K^{21}_{21})^2/K^{22}_{22}\right]} & 0\\
0 & 0 & \dfrac{i K^{22}_{31}}{1-i K_1 X^3} & 0 & 0
\end{array}\right]
+\cO(X^7).
\end{align*}
\end{widetext}

\begin{figure*}
\includegraphics[width=16cm]{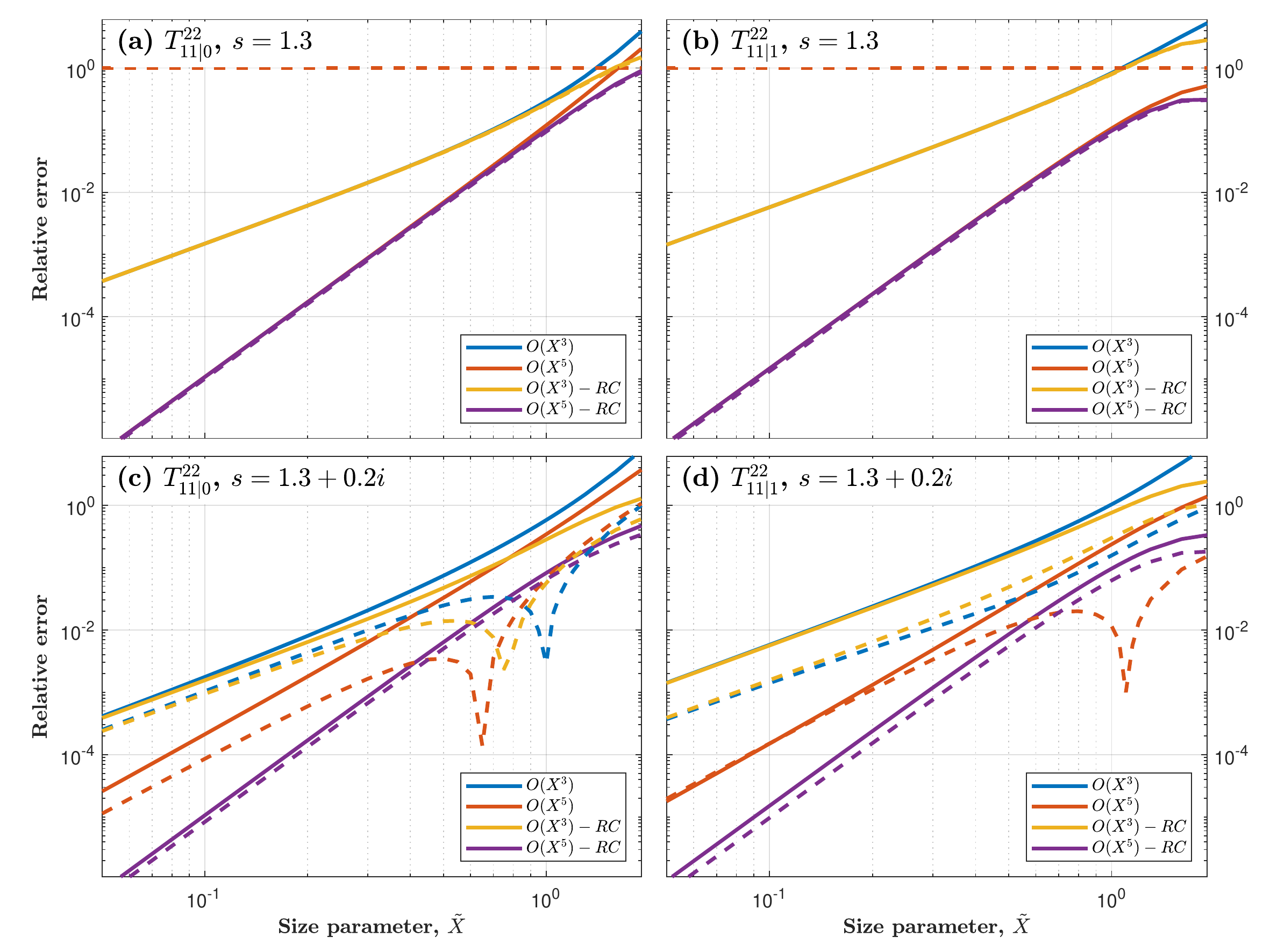}
\caption{Relative error of $T_{11|m}^{22}$ as computed to $\cO(X^3)$ or $\cO(X^5)$, with and without the radiative correction, compared to the exact solution.
The scatterer is a prolate spheroid of aspect ratio $h=3$.
We show the relative error for $|T_{11|m}^{22}|^2$ (solid lines) and $\mathrm{Re}\left[T_{11|m}^{22}\right]$ (dashed lines) for $m=0$ (left) and $m=1$ (right)
in the case of a non-absorbing material with $s=1.3$ (top) or an absorbing material with $s=1.3+0.2i$ (bottom).}
\label{FigDipolar}
\end{figure*}

\subsection{Expansions for $m\geq 2$}

The only $\cO(X^5)$ element for $m=2$ is 
\begin{align}
K_{22|2}^{22}=\frac{ X^5 }{75 h^4} \, \frac{s^2-1}{1+(s^2-1)L_{22}} + \cO(X^7)\label{K22222},
\end{align}
where we define another quadrupole depolarization factor:
\begin{align}
L_{22} 
&=\frac{1}{4e^4}\left[\frac{3}{e}(1-e^2)^2\text{atanh}(e)-3+5e^2\right].
\end{align}
Note these quadrupole factors $L_{20},L_{21},L_{22}$ follow a sum rule like the dipole factors $L_x,L_y,L_z$:
\begin{align}
L_{20}+2L_{21}+2L_{22}=2.
\end{align}
For $\b T$, the radiative correction is straightforward in this case and gives
\begin{align}
T_{22|2}^{22} = \frac{iK_{22|2}^{22}}{1-iK_{22|2}^{22}}.
\end{align}

For $m=3$, the only matrix element in our truncated blocks is $K^{22}_{33|3}$, but it is of order $\cO(X^7)$.

\subsection{Oblate spheroids and spheres}

These results were derived for prolate spheroids but also provide correct results for oblate spheroids, providing all parameters are defined in exactly the same way.
This is in contrast to other studies where different definitions are used for oblate spheroids (for example for the aspect ratio). Here, we hold:
\begin{itemize}
\item $c$ as the semi height of the spheroid along the rotation axis so that $c>a$ for prolate spheroids, $c<a$ for oblate;\\
\item $h=\dfrac{c}{a}$, so that $h>1$ for prolate, $0<h<1$ for oblate;\\
\item $e=\dfrac{\sqrt{h^2-1}}{h}$, so that $0<e<1$ for prolate, and $e$ is on the positive imaginary axis $i0<e<i\infty$ for oblate; \\  
\item $X=k_1c$ and $\tilde X=k_1 c h^{-2/3}$ in both cases. \\
\end{itemize} 

Spheres are a special case of spheroid with $h=1$ and $e=0$.
Although some expressions are singular for those values, they all have a well-defined limit.
In particular, all matrices become diagonal, $K_0 = K_1$, and
\begin{align}
\Omega_0 = \Omega_1 = \frac{3}{5}\frac{s^2-2}{s^2+2},
\end{align}
which agrees with the Taylor expansion of the electric polarizability within Mie theory \cite{2013SchebarchovPCCP}.

\section{Accuracy of new approximations}

We now assess the accuracy of these approximations by comparing them to the exact solutions, which are computed using publicly available codes, SMARTIES \cite{JQSRT2015,2016SMARTIES}, and can be obtained to an accuracy of at least $\sim 10^{-8}$. Here we use for illustration a prolate spheroid with $h=3$, but results for higher aspect ratio and for oblate spheroids are similar and included as additional figures in the supplemental material.

\subsection{Dipolar polarizability}

The dipolar polarizabilities along the $z$ and $x$ axis are proportional to $T^{22}_{11|0}$ and $T^{22}_{11|1}$ respectively.
They are the only $\cO(X^3)$ terms and therefore dominate the optical response at small size.
We compute the relative error in $|T^{22}_{11}|^2$ (related to scattering cross-section) and $\mathrm{Re}(T^{22}_{11})$ (related to extinction cross-section).
For a quantity $A^\mathrm{approx}$, the relative error $\epsilon$ is obtained by comparison with the exact result $A^\mathrm{SMARTIES}$ from:
\begin{align}
\epsilon = \left|\frac{A^\mathrm{approx}-A^\mathrm{SMARTIES}}{A^\mathrm{SMARTIES}}\right|.
\end{align}

We consider the following approximations:
\begin{itemize}
\item
$\cO(X^3)$: this includes only the dominant term and is equivalent to the Rayleigh or electrostatic approximation, i.e.
\begin{align}
T^{22}_{11|0} = iK_0 X^3,\quad T^{22}_{11|1} = iK_1 X^3.
\end{align}
\item
$\cO(X^5)$: this includes the derived $\cO(X^5)$ correction to the $K$-matrix, but not the radiative correction, so we simply have $\b T=i\b K$,
which is equivalent to a Taylor expansion of $\b T$ to order $\cO(X^5)$, i.e.
\begin{align}
T^{22}_{11|0} = \frac{iK_0 X^3}{1-\Omega_0 X^2},\quad T^{22}_{11|1} = \frac{iK_1 X^3}{1-\Omega_1 X^2}.
\end{align}
\item
$\cO(X^3)-RC$, where the radiative correction is applied to the Rayleigh approximation, i.e.
\begin{align}
T^{22}_{11|0} = \frac{iK_0 X^3}{1-iK_0 X^3},\quad T^{22}_{11|1} = \frac{iK_1 X^3}{1-iK_1 X^3}.
\end{align}
In this case, the dominant terms of $\Im(\b T)$ ($\sim X^3$) and $\Re(\b T)$ ($\sim X^6$) are both correct to lowest order. 
\item
$\cO(X^5)-RC$, where the radiative correction is applied to the $\cO(X^5)$ approximation, i.e.
\begin{align}
T^{22}_{11|0} &= \frac{iK_0 X^3}{1-\Omega_0 X^2 - iK_0 X^3},\\
T^{22}_{11|1} &= \frac{iK_1 X^3}{1-\Omega_1 X^2 -iK_1 X^3}.
\end{align}
In this case, $\b T$ is correct to order $\cO(X^6)$.
\end{itemize}

\begin{figure*}
\includegraphics[width=16cm]{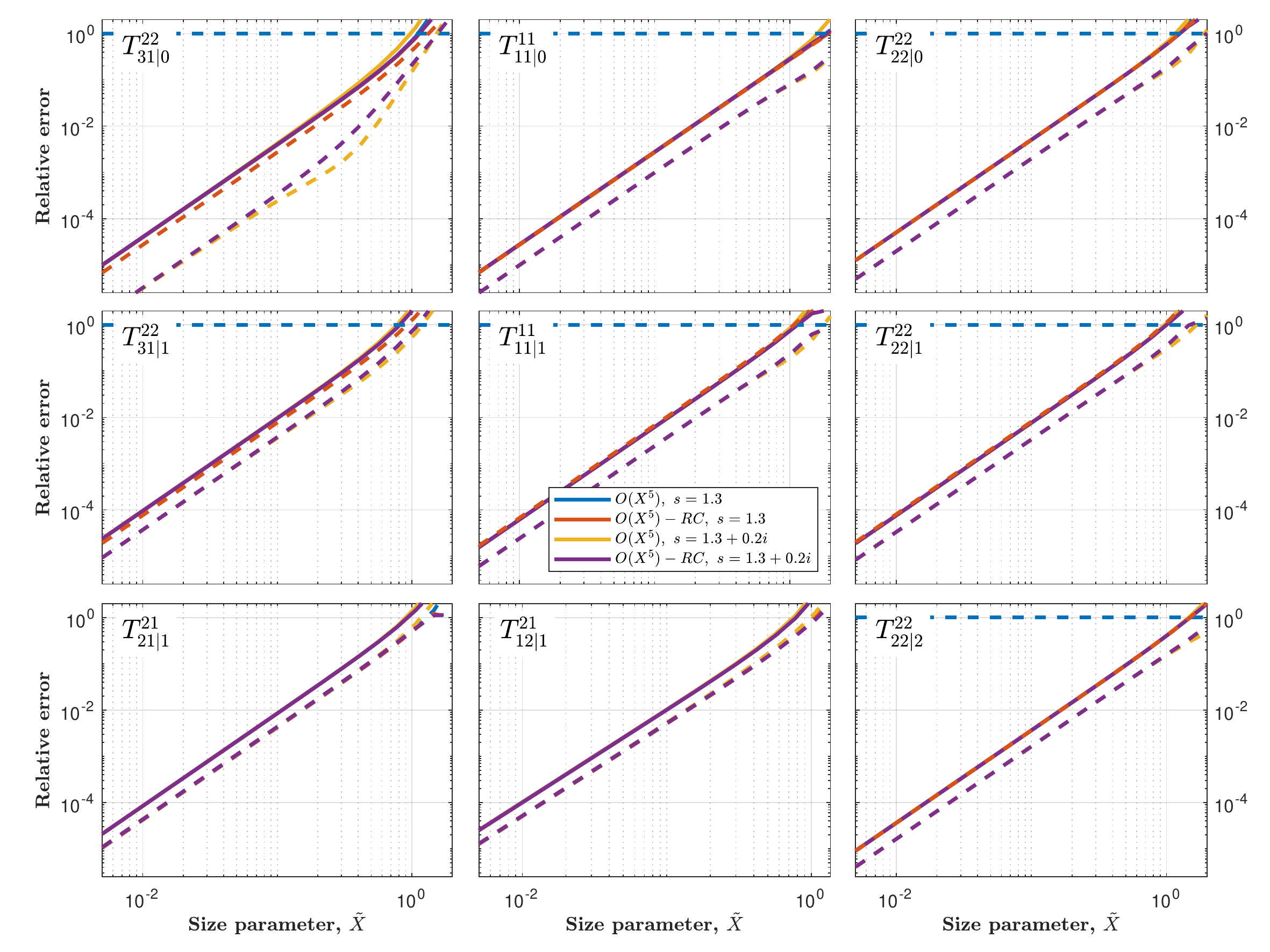}
\caption{Relative error of the other $T$-matrix elements as computed to  $\cO(X^5)$, with and without the radiative correction,
compared to the exact solution. The scatterer is a prolate spheroid of aspect ratio $h=3$.
We show the relative error for $|T_{nk|m}^{ij}|^2$ (solid lines) and $\mathrm{Re}\left[T_{nk|m}^{ij}\right]$ (dashed lines) 
in the case of a non-absorbing material with $s=1.3$ or an absorbing material with $s=1.3+0.2i$.}
\label{FigOthers}
\end{figure*}

The accuracy of these approximations as a function of the volume-equivalent size parameter $\tilde X$ is compared in Fig.~\ref{FigDipolar} for a prolate
spheroid of aspect ratio $h=3$ with either $s=1.3$ (non-absorbing) or $s=1.3+0.2i$ (absorbing). Similar
plots are provided in the supplemental material for other parameters, including higher-index ($s=1.7$) and metallic ($s=\sqrt{-10+0.5i}$) materials.
From these we can draw a number of conclusions.
\begin{itemize}
\item
Firstly, the improvements in accuracy provide reassurance that the expressions found in the previous sections are correct.
\item
Secondly, the $\cO(X^5)$ approximation clearly improves the range of validity of the approximation, as expected.
At a relative error of $10^{-2}$ for example, the $\cO(X^3)$ is applicable up to $\tilde X\approx 0.15-0.25$ while
the $\cO(X^5)$ is valid up to $\tilde X\approx 0.5-0.6$. For a particle with $h=3$ in water at a wavelength of 400\,nm, this corresponds to $c=15-25\,$nm for $\cO(X^3)$ and $c=50-60\,$nm for $\cO(X^5)$.
\item
Thirdly, the radiative correction is critical to correctly predict the real part of $T$ (and therefore the
extinction cross-section) for non-absorbing scatterers. In fact, for absorbing particles, the radiative correction
also improves further the approximation in the case of the $\cO(X^5)$-approximation. This is expected since
the radiative correction adds the correct terms up to $\cO(X^6)$. Applying the radiative correction to the Rayleigh approximation
does not improve results for absorbing scatterers as the terms of order $\cO(X^5)$ are important but not included.
\end{itemize}

In the relative error plots we can also quantify the accuracy of a particular approximation from the gradient on a log-log scale: for an approximation of the type $f(x)=C_1X^{C_2}(1+C_3X^{C_4})+\cO(X^{C_5})$, the relative error should have a gradient of $C_5-C_2$ (for small enough $X$). Hence we see the $\cO(X^3)$ approximations have a slope of 5-3=2, the $\cO(X^5)$ approximations have a slope of 6-3=3, and the $\cO(X^6)$ approximations have a slope of 7-3=4.

\begin{figure*}
\includegraphics[width=16cm]{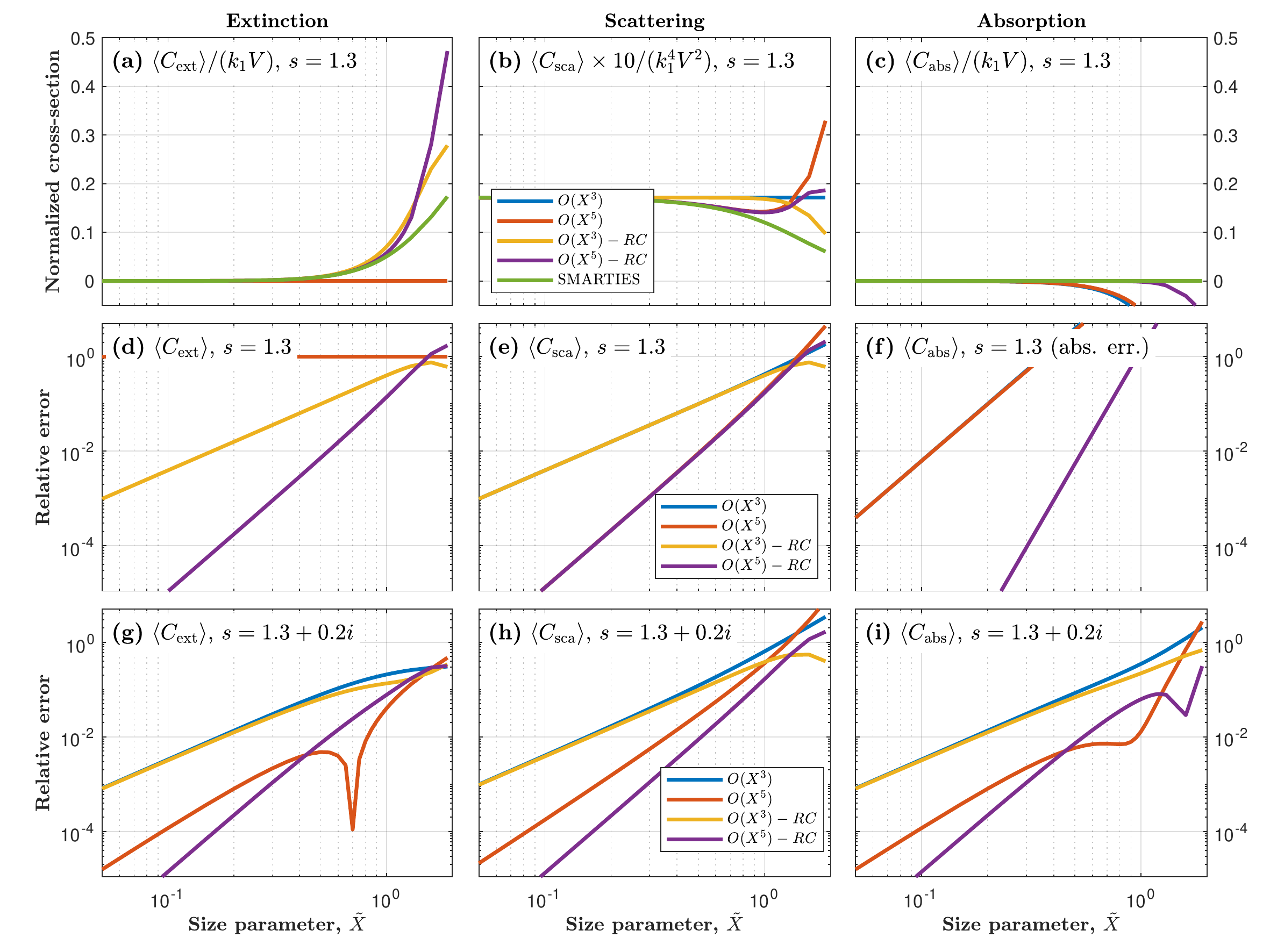}
\caption{Accuracy of approximations for predicting the orientation-averaged extinction, scattering, and absorption cross-sections.
The scatterer is a prolate spheroid of aspect ratio $h=3$, either non-absorbing with $s=1.3$ (a-f) or absorbing with $s=1.3+0.2i$ (g-i).
In the top row (a-c) the actual predicted cross-sections, suitably normalized for display, are plotted as a function of size parameter $\tilde X$
in the non-absorbing case ($s=1.3$). Unphysical negative absorption is predicted by all approximations, but for $\cO(X^5)-RC$, this only occurs for large particles outside the range of validity of the approximation. 
In (d,e,g,h,i) the relative errors (with respect to the exact results) are plotted. Since the absorption is zero in (f), we instead plot
the absolute error.}
\label{FigDerived}
\end{figure*}

\subsection{Other matrix elements}

The same tests can be carried out on the 9 other non-zero independent $T$-matrix elements, as illustrated in Fig.~\ref{FigOthers} (and comparable figures in the
supplemental material).
For all of those, the leading order is $\cO(X^5)$ and we again consider the accuracy with or without the radiative corrections,
both for $|T_{nk}^{ij}|^2$ and for $\mathrm{Re}(T_{nk}^{ij})$. These again confirm the validity of the derived expressions and lead us to similar conclusions
as those obtained for the dipolar polarizability.
In these other matrix elements, the radiative correction is of order $X^{10}$, and does not improve much the approximation for absorbing scatterers
(unlike for the dipolar polarizability). The radiative correction does remain crucial to avoid zero-extinction and negative absorption in the case
of non-absorbing scatterers.

\subsection{Derived quantities}
 

One of the main advantages of the $T$-matrix framework is the possibility to compute efficiently orientation-averaged properties \cite{2002Mishchenko}.
The orientation-averaged extinction cross-section is obtained from the trace of $\b T$, so up to order $X^6$ only depends on $T^{11}_{11|m}$, $T^{22}_{11|m}$, and $T^{22}_{22|m}$.
Explicitly:
\begin{align}
&\langle C_\text{ext}\rangle 
=\frac{-2\pi}{k_1^2} \Re\big\{ T_{11|0}^{11} + T_{11|0}^{22} + T_{22|0}^{22} \nonumber\\
&+  2[ T_{11|1}^{11} + T_{11|1}^{22} + T_{22|1}^{22} + T_{22|2}^{22} ] \big\} + \cO(X^7). \label{Cext OX5}
\end{align}

The orientation-averaged scattering cross section is computed from the sum of the squares of the elements of $\b T$.
We have 11 independent non-zero matrix elements to $\cO(X^6)$ and therefore: 
\begin{align}
\langle C_\text{sca} \rangle &= \frac{2\pi}{k_1^2} 
\big\{ |T_{11|0}^{11}|^2 + |T_{11|0}^{22}|^2 + |T_{22|0}^{22}|^2 + 2|T_{31|0}^{22}|^2 + \nonumber\\
& 2|T_{11|1}^{11}|^2 + 2|T_{11|1}^{22}|^2 + 4|T_{12|1}^{12}|^2 + 4|T_{21|1}^{21}|^2 + \nonumber\\[0.2cm]
& 4|T_{31|1}^{22}|^2 + 2|T_{22|1}^{22}|^2 + 2|T_{22|2}^{22}|^2 \big\} + \cO(X^7). \label{Csca OX5}
\end{align}

The accuracy of the new approximations for the orientation-averaged cross-sections is illustrated in Fig.~\ref{FigDerived} (and comparable figures in the supplemental material), including and not including the radiative corrections. These follow the same trends as for the individual matrix elements.

\section{Conclusion}

We have obtained Taylor expansion in the long-wavelength or low-frequency limit of all the elements of the $T$-matrix for prolate and oblate spheroidal particles up
to order $\cO(\tilde X^6)$, where $\tilde X$ is the size parameter. The coefficients of these expansions are simple expressions in terms of the relative refractive index $s$ and aspect ratio $h$. The resulting approximation expands the range of validity of the commonly-used Rayleigh approximation from size parameters of the order of $\tilde X\approx 0.2$ up to $\tilde X\approx 0.6$ depending on the parameters.
For a particle in water of aspect ratio $h=3$ at a wavelength of 400\,nm, this corresponds to maximum dimensions of $40\,$nm for the Rayleigh approximation, increasing to $120\,$nm for our new approximations.
This new approximation is therefore applicable to a much wider range of nanoparticles commonly synthesized and studied, for which the
Rayleigh approximation is typically inadequate.
In addition, this study provides further insight into the importance of the radiative correction and the related $K$-matrix \cite{2013LeRuPRA}
and how it can be used effectively to obtain more accurate and physical results with approximations of the $T$-matrix.
This work is intended as a simple alternative to the full $T$-matrix calculations to further study the optical properties of nanoparticles.

\begin{acknowledgments}
The authors thank the Royal Society Te Ap\=arangi (New Zealand) for support through a Rutherford Discovery Fellowship (B.A.)
and a Marsden grant (E.C.L.R.).
\end{acknowledgments}

\appendix

{\large \bf Appendix}

\section{Integral forms of matrix elements}
\label{AppIntegrals}

The elements of the $P$-, $Q$- and $U$- matrices can be computed from integrals of Bessel functions. The only difference is the type of Riccati-Bessel functions used, $\psi_n$ for $\b P$, $\xi_n$ for $\b Q$, and $\chi_n$ for $\b U$:
\begin{align}
\psi_n(x)&=xj_n(x), \quad\chi_n(x) = x y_n(x),\nonumber \\
\xi_n(x)&=x h_n(x)=\psi_n(x)+i\chi_n(x),
\end{align}
where $j_n$, $y_n$ and $h_n$ are the spherical Bessel and Hankel functions.

Expressions below are given for $\b Q$ only, as the others are easily derived by replacing $\xi_n$. 
Here we recap the simplified expressions obtained in \cite{2011SomervilleOL} and provide new expressions for the diagonal of $\b Q$.
The letter $K$ and $L$ are used for the integrals to comply with previous work \cite{2011SomervilleOL}, but
should not be confused with the $K$-matrix elements or the depolarization factors also denoted $L$.
Note that these integrals are exact for axisymmetric particles of any size. Below we assume $\xi_n=\xi_n(x), \psi_n=\psi_n(sx)$ to shorten expressions,
and the prime denotes the derivative.
\begin{align}
Q^{12}_{nk}&=A_nA_k\frac{s^2-1}{s} K^1_{nk}, \label{EqnQ12simple}\\
Q^{21}_{nk}&=A_nA_k\frac{1-s^2}{s} K^2_{nk}, \label{EqnQ21simple}\\
Q^{11}_{nk}&= -iA_nA_k\left[- s L^1_{nk} + L^3_{nk}+\frac{L^2_{nk} - L^4_{nk}}{s} \right],\\
Q^{22}_{nk}&= -iA_nA_k\left[- L^1_{nk} + \frac{L^3_{nk}}{s}+L^2_{nk} - L^4_{nk} \right],  \label{EqQ22} 
\end{align}
where
\begin{equation}
A_n = \sqrt{\frac{2n+1}{2n(n+1)}},
\end{equation}
\begin{align}
K^1_{nk}=&m\int^\pi_0 d\theta  d_{n} d_{k} \xi_n\psi^\prime_kx_\theta ,\label{K1}\\
K^2_{nk}=&m\int^\pi_0 d\theta  d_{n} d_{k}  \xi^\prime_n\psi_kx_\theta ,\\
L^1_{nk}= &\int^\pi_0
d\theta \sin\theta \tau_{n}d_{k}\xi_n\psi_kx_\theta ,\\
L^2_{nk} = &\int^\pi_0
d\theta\sin\theta d_{n}\tau_{k}\xi_n\psi_kx_\theta ,
\end{align}
\begin{align}
L^3_{nk} &= \int^\pi_0
d\theta \sin\theta d_{k} \psi^\prime_k \left[x_\theta \tau_{n} \xi^\prime_n  -n(n+1)d_{n} \xi_n\right] \nonumber\\
 &\equiv L^{31}_{nk}+L^{32}_{nk},\\
L^4_{nk} &= \int^\pi_0
d\theta \sin\theta d_{n} \xi^\prime_n \left[sx_\theta \tau_{k} \psi^\prime_k - k(k+1)d_{k} \psi_k(sx)\right] \nonumber\\
&\equiv L^{41}_{nk}+L^{42}_{nk}.\label{L4}
\end{align}
$d_{n}$ and $\tau_{n}$ are angular functions (related to Wigner's $d$-function) \cite{2002Mishchenko,2011SomervilleOL} defined in terms of
the associated Legendre functions $P_{n}^{m}$:
\begin{align}
&P_n^m(\cos\theta)=\sin^m\theta \left(\frac{\d}{\d\cos\theta}\right)^mP_n(\cos\theta) \\
&d_{n} \! =\! (-)^m\sqrt{\frac{(n-m)!}{(n+m)!}}P_{n}^{m}(\cos\theta), ~ \pi_n\!=\!\frac{md_n}{\sin\theta},~
\tau_{n} \!=\! \frac{\d}{\d\theta}d_{n}.
\end{align}

For {\it off-diagonal elements} ($n \neq k$), the four $L^i$ integrals are not linearly independent and we may therefore
also use the following simplifications:
\begin{align}
Q^{11}_{nk}=& iA_nA_k\frac{s^2-1}{s}
 \frac{n(n+1)L^2_{nk}-k(k+1)L^1_{nk}}{n(n+1)-k(k+1)}, \label{EqnQ11simple}\\
Q^{22}_{nk}=& iA_nA_k\frac{s^2-1}{s}
\left[L^3_{nk}+\frac{sn(n+1)(L^2_{nk}-L^1_{nk})}{n(n+1)-k(k+1)}\right] \label{EqnQ22simple}.
\end{align}

For {\it diagonal elements} ($n = k$), we moreover have:
\begin{align}
L^1_{nn}=L^2_{nn}, \qquad sL^{31}_{nn}=L^{41}_{nn},
\end{align}
and we therefore instead use:
\begin{align}
Q^{11}_{nn}= L^{51}_n +iA_nA_n\frac{s^2-1}{s}L^1_{nn},\\
Q^{22}_{nn}= L^{52}_n +iA_nA_n\frac{s^2-1}{s} L^{31}_{nn},
\end{align}
where
\begin{align}
L^{51}_{n}&=\frac{-iA_nA_n}{s}(sL^{32}_{nn}-L^{42}_{nn}) \\
&=  -\frac{i}{s}\frac{2n+1}{2} \int^\pi_0
d\theta \sin\theta d_{n} d_{n}\left[ \xi^\prime_n \psi_n - s \xi_n \psi^\prime_n\right ], \nonumber\\[0.5cm]
L^{52}_{n}&=-i\frac{A_nA_n}{s} (L^{32}_{nn}-sL^{42}_{nn}) \\
 &= -\frac{i}{s}\frac{2n+1}{2} \int^\pi_0
d\theta \sin\theta d_{n} d_{n}\left[ s\xi^\prime_n \psi_n - \xi_n \psi^\prime_n\right ].\nonumber
\end{align}
These expressions are convenient because:
\begin{itemize}
\item
For $s=1$, we have $L^{51}_{n}=L^{52}_{n}=1$, which implies that $\b Q$ is the identity as desired
(all the other terms contain a $s^2-1$ factor).
\item
$L^{51}_{n}$ and $L^{52}_{n}$ are analogous to the terms appearing in Mie theory, and in some sense a generalization of these.
\end{itemize}

We have derived and used here an alternative form with similar features.
On can show using integration by parts that:
\begin{align}
&sL^{3}_{nn}-L^{4}_{nn} = \\
&\int^\pi_0 d\theta \sin\theta (\pi_n\pi_n+\tau_n\tau_n)\left[ \xi^\prime_n \psi_n - s \xi_n \psi^\prime_n\right ]+ (s^2-1) L^1_{nn}, \nonumber
\end{align}
and
\begin{align}
&L^{3}_{nn}-sL^{4}_{nn} = \\
&\int^\pi_0 d\theta \sin\theta (\pi_n\pi_n+\tau_n\tau_n)\left[ s\xi^\prime_n \psi_n - \xi_n \psi^\prime_n\right ]+ \frac{s^2-1}{s} L^7_{nn}, \nonumber
\end{align}
where
\begin{align}
L^7_{nn}= n(n+1)\int^\pi_0
d\theta \sin\theta \tau_{n}d_{n} x_\theta \frac{\xi_n\psi_n}{x^2}.
\end{align}
If we also define:
\begin{align}
L^{61}_{n} &= \frac{i}{s}\frac{2n+1}{2} \int^\pi_0
d\theta \sin\theta (\pi_{n} \pi_{n}+\tau_n\tau_n)\left[ s \xi_n \psi^\prime_n - \xi^\prime_n \psi_n\right ],\\
L^{62}_{n} &= \frac{i}{s}\frac{2n+1}{2} \int^\pi_0
d\theta \sin\theta (\pi_{n} \pi_{n}+\tau_n\tau_n)\left[ \xi_n \psi^\prime_n - s\xi^\prime_n \psi_n\right ],
\end{align}
we then have:
\begin{align}
Q^{11}_{nn}&=   L^{61}_n, \\
Q^{22}_{nn}&= L^{62}_n -iA_nA_n\frac{s^2-1}{s} L^{7}_{nn}.
\end{align}

These expressions are used in the next section to determine the asymptotic behaviour of
the matrix elements for low size.

\section{Size dependence of matrix elements}

\subsection{Integrals for $\b P$, $\b Q$, and $\b U$}
\label{AppIntOrder}
We investigate the lowest orders in $X$ for the $P$- and $Q$- matrices in order to determine where to truncate them when computing the $T$-matrix.
We can expand the Bessel functions in powers of $x$ to obtain analytic expressions for the low orders.
At lowest order, we can use $\xi_n \sim x^{-n}$ and $\psi_n\sim x^{n+1}$.
The hat above the integral letters below will be used when it applies to the $P$-matrix (i.e. $\xi_n$ replaced by $\psi_n$).
The $U$-matrix behaves asymptotically like the $Q$-matrix.
We have:
\begin{itemize}
\item
$K^1_{nk}, K^2_{nk} \propto X^{-n+k+1}$. Determines $Q^{12}_{nk},Q^{21}_{nk}$.
\item
$\hat{K}^1_{nk}, \hat{K}^2_{nk} \propto X^{n+k+2}$. Determines $P^{12}_{nk},P^{21}_{nk}$.
\item
$L^1_{nk}, L^2_{nk} \propto X^{-n+k+2}$. Determines $Q^{11}_{nk}$ for $n \neq k$.
\item
$\hat{L}^1_{nk}, \hat{L}^2_{nk} \propto X^{n+k+3}$. Determines $P^{11}_{nk}$ for $n \neq k$.
\item
$L^3_{nk}, L^4_{nk} \propto X^{-n+k}$. Determines $Q^{22}_{nk}$ for $n \neq k$.
\item
$\hat{L}^3_{nk}, \hat{L}^4_{nk} \propto X^{n+k+1}$. Determines $P^{22}_{nk}$ for $n \neq k$.
\item
$L^{61}_{n}, L^{62}_{n}, L^{7}_{nn} \propto X^0$. Determines $Q^{11}_{nn}$ and $Q^{22}_{nn}$.
\item
$\hat{L}^{61}_{n} \propto X^{2n+3}$ (note this is a special case as the highest order terms cancel). Determines $P^{11}_{nn}$.
\item
$\hat{L}^{62}_{n}, \hat{L}^{7}_{nn} \propto X^{2n+1}$. Determines $P^{22}_{nn}$.
\end{itemize}

These justify the small size behaviour of all matrix elements given in the main text.

\subsection{Proof of lowest-order $X$-dependence for $\b T$ and $\b R$ matrices}
\label{AppInvGen}

In order to determine the orders of $\b T$, we must first determine the orders of $\b R$ by inverting $\b Q$. 
For this we use the blockwise matrix inversion formula:
\begin{align}
& \vc{ \b R^{11} & \b R^{12} \\
\b R^{21} & \b R^{22} } 
=  \vc{ \b Q^{11} & \b Q^{12} \\
\b Q^{21} & \b Q^{22} }^{-1} \label{R blockwise} \\[0.5cm]
 \mathrm{where}&\nonumber\\
\b R^{11} &= ( \b Q^{11} - \b Q^{12} [\b Q^{22}]^{-1}\b Q^{21} )^{-1} \nonumber \\
\b R^{12} &= -\b R^{11} \b Q^{12} [\b Q^{22}]^{-1} \nonumber\\
\b R^{21} &= -[\b Q^{22}]^{-1} \b Q^{21} \b R^{11} \nonumber\\ 
\b R^{22} &= [\b Q^{22}]^{-1} + [\b Q^{22}]^{-1} \b Q^{21} \b R^{11} \b Q^{12} [\b Q^{22}]^{-1}\nonumber 
\end{align}

We first focus on general axisymmetric particles for which the $X$-dependence of $\b P$ and $\b Q$ was derived.
In order to use \eqref{R blockwise}, we first need to 
determine the order of $[\b Q^{22}]^{-1}$. This can be done by induction on the matrix truncation order $N$.
For this proof our inductive assumption is that for a general $N\times N$ matrix $\b M$
with $M_{nk}\propto X^{k-n}$, then $[M^{-1}]_{nk}\propto X^{k-n}$.
The base case ($N=1$) is trivial. For the inductive step, we use the blockwise decomposition, but with $N\times N$, $N\times1$, $1\times N$ and $1\times1$ blocks.

\begin{align*}
&[\b M]^{-1}(N+1\times N+1) = 
\left[ \begin{array}{c|c}
\\~\quad \b A ~\quad 
& \b B \\ \\ \hline
~\quad \b C ~\quad 
& D 
\end{array} \right]^{-1} \\
&=
\left[ \begin{array}{c|c}
 \b E \quad & -\b E \b B D^{-1} \\ \hline
-D^{-1} \b C\b E & D^{-1} + D^{-1}\b C\b E\b B D^{-1}  
\end{array} \right] \\[0.5cm]
&\mathrm{where~}\b E = [\b A - \b B D^{-1} \b C ]^{-1}.
\end{align*}
We need to show that
\begin{align}
E_{nk} &\propto A_{nk} \propto X^{k-n} \label{prop proof1}\\
[-\b E\b B D^{-1}]_n &\propto B_n \propto X^{N+1-n} \label{prop proof2}\\
[-D^{-1} \b C\b E]_k &\propto C_k \propto X^{k-N-1} \label{prop proof3}\\
D^{-1} + D^{-1}\b C\b E\b B D^{-1} &\propto D \propto X^0. \label{prop proof4}
\end{align}
These can easily be proved from the definition of matrix product.
For example \eqref{prop proof1} follows from
\begin{align}
A_{nk} - [\b B D^{-1} \b C]_{nk} = A_{nk} - D^{-1}B_nC_k \propto X^{k-n}.
\end{align}
The rest (\ref{prop proof2}-\ref{prop proof4}) follow from similar straightforward derivations. This then proves the inductive assumption.

To find the orders of $\b R$, we will also require to know the inverse for matrices with $M_{nk}\propto X^{k-n+2-2\delta_{nk}}$, which can also be proven by induction to be $\left[M^{-1}\right]_{nk}\propto X^{k-n+2-2\delta_{nk}}$. 

Then for $\b R^{11}$ we have
\begin{align}
[\b Q^{12}[\b Q^{22}]^{-1}\b Q^{21}]_{nk} &= \sum_{p=1}^N\sum_{q=1}^N Q^{12}_{np} [Q^{22}]^{-1}_{pq} Q^{21}_{qk} \\
&\propto \sum_{p,q} \ldots ~ X^{p-n+1}X^{q-p}X^{k-q+1} \\
&\propto X^{k-n+2}
\end{align}
None of these orders dominate the elements in $\b Q^{11}$, so we have $[ \b Q^{11} - \b Q^{12} [\b Q^{22}]^{-1}\b Q^{21} ]_{nk} \propto X^{k-n+2-2\delta_{nk}} \Rightarrow R^{11}_{nk}\propto X^{k-n+2-2\delta_{nk}}$.
The derivations for the other blocks in $\b R$ are similar and it turns out that the orders are exactly the same as for $\b Q$.  
Then for $\b T$:
\begin{align}
\vc{ \b T^{11} & \b T^{12} \\
\b T^{21} & \b T^{22} } = \vc{ 
\b P^{11}\b R^{11} + \b P^{12}\b R^{21} & 
\b P^{11}\b R^{12} + \b P^{12}\b R^{22} \\
\b P^{21}\b R^{11} + \b P^{22}\b R^{21} & 
\b P^{21}\b R^{12} + \b P^{22}\b R^{22} }
\end{align}
and again the orders of $\b T$ can be found from straightforward matrix multiplication to be identical to those of $\b P$.

For particles with mirror symmetry, the orders for $\b P$ and $\b Q$ are the same as in the general case, but half the elements are zero, as discussed in section \ref{SecMirror}. The same derivation would show that the orders for $\b R$ and $\b T$ are the same as for the general case except for the zero-elements which are zero by symmetry.

\subsection{Special case of spheroids}
\label{AppInvSph}

For spheroids we have to modify the derivation to account for the elements of $\b Q$ below the diagonal of each block, which are all $\cO(X^0)$
as shown in \cite{JQSRT2012}. This may be expressed as
\begin{align}
Q^{11}_{nk} &\propto X^{[n<k](k-n+2)}, \quad Q^{22}_{nk} \propto X^{[n<k](k-n)}, \nonumber\\
Q^{12}_{nk} &\propto X^{[n<k](k-n+1)}, \quad Q^{21}_{nk} \propto X^{[n<k](k-n+1)},
\end{align}
where $[n<k] = 1$ if $n<k$ and 0 otherwise. Note that half of those matrix elements are zero, but this does not affect the derivation.
To find the order of $\b R^{21}$ for example, we need the product
\begin{align}
[[\b Q^{22}]^{-1}\b Q^{21}]_{nk}\! \propto \!\sum_{p=1}^N\! \ldots X^{[n<p](p-n)}X^{[p<k](k-p+1)}
\end{align}
which can be broken down into two cases. Firstly for $n<k$, the sum may contain terms with $p\leq n<k$ resulting in order $X^{k-p+1}$,
or $n<p<k$ giving $X^{k-n+1}$, or $n<k\leq p$ giving $X^{p-n}$. From this we simply take the dominant order: $X^{k-n+1}$.
Secondly, for $n\geq k$ it can be shown using similar arguments that $[[\b Q^{22}]^{-1}\b Q^{21}]_{nk}\propto X^0$.
Following through with similar derivations for all the other terms, we find again that $\b R$ behaves asymptotically like $\b Q$ and $\b T$ like $\b P$.

\section{Taylor expansions of matrix elements}

\subsection{Multipole truncation and matrix inversion}
\label{AppInvTrunc}

The first step in determining the Taylor expansions is to find the truncation needed to reach a specific
accuracy. For a matrix product, it is relatively easy to track the orders.
To compute the matrix $\b K^{eo} = \b P^{eo} [\b U^{eo}]^{-1}$ to order $X^5$, we have seen in the main text that
both $\b P^{eo}$ and $[\b U^{eo}]^{-1}$ can be truncated at multipole $N=3$.
But when carrying the matrix inversion of $\b U^{eo}$, any matrix element of the inverse depends in principle in a non-trivial way on all the other matrix elements.
Truncating at multipole $N=5$ and writing out leading orders explicitly, we have for spheroidal particles:
\begin{align}
\mathbf{U}^{eo}= 
\left[\def\arraystretch{1.4}\begin{array}{cc|ccc}
u^{11}_{22}X^{ 0} & u^{11}_{24}X^{ 4} & u^{12}_{21}X^{ 0} & u^{12}_{23}X^{ 2} & u^{12}_{25}X^{ 4 }\\
u^{11}_{42}X^{ 0} & u^{11}_{44}X^{ 0} & u^{12}_{41}X^{ 0} & u^{12}_{43}X^{ 0} & u^{12}_{45}X^{ 2} \\
\hline
u^{21}_{12}X^{ 2} & u^{21}_{14}X^{ 4} & u^{22}_{11}X^{ 0} & u^{22}_{13}X^{ 2} & u^{22}_{15}X^{ 4} \\
u^{21}_{32}X^{ 0} & u^{21}_{34}X^{ 2} & u^{22}_{31}X^{ 0} & u^{22}_{33}X^{ 0} & u^{22}_{35}X^{ 2} \\
u^{21}_{52}X^{ 0} & u^{21}_{54}X^{ 0} & u^{22}_{51}X^{ 0} & u^{22}_{53}X^{ 0} & u^{22}_{55}X^{ 0}
\end{array}\right],
\end{align}
where the coefficients $u^{ij}_{nk}$ are all of order $X^0$.
We have shown in the previous section that the inverse, whose matrix elements are denoted $v^{ij}_{nk}$, has the same leading orders.
By carrying the inversion explicitly using the inversion formula with the matrix of co-factors (or using a symbolic calculation software),
one find that the leading orders of the 9 elements $v^{ij}_{nk}$ associated with a multipolarity of 3 or less (i.e. with $1\le n,k \le 3$)
only depend on the $u^{ij}_{nk}$ with a multipolarity of 3 or less (i.e. with $1\le n,k \le 3$).
In order to obtain $\b K^{eo}$ to order 5, we moreover need to expand $u^{22}_{11}$ to the next order ($X^2$).
Again, we can show that this next order correction only depends on terms of multipolarity of 3 or less.

A similar analysis can be carried out for the $oe$ matrix, where it can be concluded that truncation at multipolarity $N=2$ is sufficient to compute $\b K^{oe}$ to order $X^5$.
We reiterate that this is a special property of spheroidal particles and would not apply to another general shape.

\subsection{Accuracy to $\mathcal{O}(X^{5})$ - inverting $\b U^{eo}$}
\label{AppInvU}

Following up from the previous section, we now calculate explicitly $\b K^{eo}$ to order $X^5$.
Truncating all matrices at multipolarity $N=3$ and tracking orders during matrix inversion and multiplication, one can
show that the only relevant terms in the expansions for $\b U^{eo}$ and $\b P^{eo}$ are:
\begin{align} 
\b U^{eo}  =  
\begin{bmatrix}
\def\arraystretch{1.4}\begin{array}{c|cc}
u_{22}   & u_{21}  & u_{23}X^{2}  \\
\hline
u_{12}X^{2}  & u_{11} + u_{11}^{(2)}X^{2} & u_{13}X^{2}  \\
u_{32}  & u_{31}  & u_{33}   
\end{array}
\end{bmatrix},
\end{align}
\begin{align}
\b P^{eo} = 
\begin{bmatrix}
\def\arraystretch{1.4}\begin{array}{c|cc}
0 & p_{21}X^{5} &  0\\
\hline
p_{12}X^{5} &  p_{11}X^{3} + p_{11}^{(2)}X^{5}  & p_{13}X^{5} \\
 0 &p_{31}X^{5} & 0
\end{array}
\end{bmatrix}
\end{align}
where we have omitted the superscripts in $u_{nk}$ and $p_{nk}$. These coefficients, which are of order $X^0$ by construction, can be calculated by substituting the appropriate power series for the integrands of $K$ and $L$ (or $\hat{K}$ and $\hat{L}$) integrals.
$[\b U^{eo}]^{-1}$ is obtained by direct inversion as:
\begin{widetext}
\begin{align} 
[\b U^{eo}]^{-1}  =  
\begin{bmatrix}
\def\arraystretch{2.2}\begin{array}{c|cc}
\frac{1}{u_{22}}X^0 +\cO(X^2) & -\frac{u_{21}}{u_{11}u_{22}} X^0 +\cO(X^2)  & -\frac{u_{11}u_{23} - u_{13}u_{21}}{u_{11}u_{22}u_{33}}X^2+ \cO(X^4)  \\
\hline
-\frac{u_{12}u_{33} - u_{13}u_{32}}{u_{11}u_{22}u_{33}}X^2+ \cO(X^4) & \frac{1}{u_{11}} \frac{X^0}{1-\Omega_u X^2+\cO(X^4) } & - \frac{u_{13}}{u_{11}u_{33}}X^2+ \cO(X^4)   \\
-\frac{u_{32}}{u_{22}u_{33}} X^0 +\cO(X^2) & \frac{u_{21}u_{32} - u_{22}u_{31}}{u_{11}u_{22}u_{33}} X^0 +\cO(X^2)  & \frac{1}{u_{33}}  X^0 +\cO(X^2) 
\end{array}
\end{bmatrix},
\end{align}
where
\begin{align}
\Omega_u = \frac{u_{13}u_{21}u_{32} - u_{13}u_{22}u_{31} -u_{12}u_{21}u_{33}  + u_{11}^{(2)}u_{22}u_{33}}{u_{11}u_{22}u_{33}}.
\end{align}
Note that we used $1 + ax^{2} + \cO(x^4) = (1-ax^{2} + \cO(x^4))^{-1}$ for reasons that are explained in the text. 

Carrying out the matrix multiplication $\b K^{eo} =  \b P^{eo} [\b U^{eo}]^{-1}$ and keeping only terms of order $X^5$ or less, we get
\begin{align} 
\b K^{eo} =  
\begin{bmatrix}
\def\arraystretch{2.2}\begin{array}{c|cc}
0 & \frac{p_{21}}{u_{11}} X^5  & 0\\
\hline
\frac{u_{11}u_{33}p_{12} + u_{13}u_{32}p_{11} -u_{11}u_{32}p_{13} - u_{12}u_{33}p_{11}}{u_{11}u_{22}u_{33}}X^5 & \frac{p_{11}}{u_{11}}X^{3} \frac{1}{1 - \Omega X^{2} + \cO(X^{4})} & \frac{u_{11}p_{13} - u_{13}p_{11}}{u_{11}u_{33}} X^5 \\
0 & \frac{p_{31}}{u_{11}} X^5   & 0 
\end{array}
\end{bmatrix}+\cO(X^7),
\end{align}
where
\begin{equation} \label{eqn:Omega}
\Omega = \frac{p_{11}^{(2)}}{p_{11}} - \frac{u_{11}^{(2)}}{u_{11}} + \frac{u_{21}}{u_{22}} \left( \frac{u_{12}}{u_{11}} -  \frac{p_{12}}{p_{11}} \right)- \left( \frac{u_{21}}{u_{22}} \frac{u_{32}}{u_{33}} - \frac{u_{31}}{u_{33}}\right) \left( \frac{u_{13}}{u_{11}} - \frac{p_{13}}{p_{11}} \right).
\end{equation}
\end{widetext}
When $m=0$, all $u_{nk}$ and $p_{nk}$ with $n+k$ odd are also zero as they belong to the off-diagonal blocks, which yields a much simpler expression for $\Omega$. 
All these expressions can be further simplified once these coefficients have been calculated explicitly.

\subsection{Accuracy to $\mathcal{O}(X^{5})$ - inverting $\b U^{oe}$}

A similar analysis can be carried out for $\b U^{oe}$. Expressions are somewhat simpler since the matrix
can be truncated at multipolarity $N=2$.
Also, the only relevant terms in the expansions for $\b U^{oe}$ and $\b P^{oe}$ are the dominant terms:
\begin{align} 
\b U^{oe}  =  
\begin{bmatrix}
\def\arraystretch{1.4}\begin{array}{c|c}
u^{11}_{11}   & u^{12}_{12} X^2  \\
\hline
u^{21}_{21}   & u^{22}_{22}
\end{array}
\end{bmatrix},
\quad
\b P^{oe} = 
\begin{bmatrix}
\def\arraystretch{1.4}\begin{array}{c|c}
p^{11}_{11} X^5   & p^{12}_{12} X^5  \\
\hline
p^{21}_{21} X^5  & p^{22}_{22} X^5
\end{array}
\end{bmatrix}.
\end{align}
The block superscripts are here written explicitly as the coefficients are different
to those defined in the $eo$ case.

Carrying out the matrix inversion, we have:
\begin{align} 
[\b U^{oe}]^{-1}  =  
\begin{bmatrix}
\def\arraystretch{2.2}\begin{array}{c|c}
\frac{1}{u^{11}_{11}} & 0\\
\hline
-\frac{u^{21}_{21}}{u^{11}_{11}u^{22}_{22}} & \frac{1}{u^{22}_{22}}
\end{array}
\end{bmatrix}+\cO(X^2),
\end{align}
from which we deduce
\begin{align} 
\b K^{oe}  =  
\begin{bmatrix}
\def\arraystretch{2.2}\begin{array}{c|c}
\left(\frac{p^{11}_{11}}{u^{11}_{11}} -\frac{p^{12}_{12}u^{21}_{21}}{u^{11}_{11}u^{22}_{22}}\right)X^5 
 & \frac{p^{12}_{12}}{u^{22}_{22}}X^5  \\
\hline
\left(\frac{p^{21}_{21}}{u^{11}_{11}} -\frac{p^{22}_{22}u^{21}_{21}}{u^{11}_{11}u^{22}_{22}}\right)X^5 
& \frac{p^{22}_{22}}{u^{22}_{22}}X^5
\end{array}
\end{bmatrix}+ \cO(X^7).
\end{align}

\subsection{Calculating expansion coefficients for the integrals}

To demonstrate the method for calculating the Taylor expansion of the matrix elements from the integrals, we use $U_{11|0}^{22}$ as an example and compute the first two orders, $u_{11}$ and $u_{11}^{(2)}$. The exact expression we choose to start from is
\begin{align}
U^{22}_{11}&=L^{52}_1+\frac{3i}{4}\frac{s^2-1}{s}L^{31}_{11}
\end{align}
where
\begin{align}
L^{31}_{11} &= \int^\pi_0
\d\theta \sin\theta d_{1}\tau_{1} \chi_1' \psi^\prime_1    x_\theta, \\
L^{52}_{1} &= \frac{3i}{2s} \int^\pi_0 \d\theta \sin\theta d_{1}d_1\left[ \chi_1(x) \psi^\prime_1(sx) - s\chi^\prime_1(x) \psi_1(sx)\right ].
\end{align}
The Bessel functions in the integrand have the following Taylor expansions:
\begin{align*}
\chi_1'(x)\psi_1'(sx)&=\frac{2s}{3}x^{-1} - \frac{s(2s^2+5)}{15}x +\mathcal{O}(x^3),
\end{align*}
\begin{align*}
s\chi^\prime_1(x)&\psi_1(sx)-\chi_1(x)\psi^\prime_1(sx) = \\
&\frac{1}{3}s(s^2+2) - \frac{1}{30}s(s^2-1)(s^2+10)x^2 +\mathcal{O}(x^4).
\end{align*}
Inserting these we have
\begin{align*}
L_{11}^{31}&= -\frac{2s}{3}\int_0^\pi \d\theta \cos\theta \sin^2\theta  x_\theta x^{-1} + \\
& \qquad\frac{s(2s^2+5)}{15} \int_0^\pi \d\theta \cos\theta \sin^2\theta   x_\theta x + \mathcal{O}(X^4),\\[0.5cm]
L^{52}_{1} &= -i\frac{s^2+2}{3} + \\
&\qquad i\frac{(s^2-1)(s^2+10)}{20}\int_0^\pi \d\theta\! \cos^2\theta \sin\theta x^2   + \mathcal{O}(X^4). 
\end{align*}
Putting these together and rearranging, we obtain expressions for $u_{11}$ and $u_{11}^{(2)}$ for $m=0$ in terms of angular
integrals involving $x(\theta)$ and $x_\theta(\theta)$ and other angular functions.

We found that in general the angular integrals appearing in the $u_{nk}$ coefficients can be expressed in terms of Legendre polynomials
of the second kind $Q_n^m\equiv Q_n^m(\xi_0)$, where $\xi=\xi_0=1/e=h/\sqrt{h^2-1}$ is the spheroidal coordinate defining the surface of the spheroid. 
These integrals are also related to the depolarization factors, for example $L_z$ can be written as 
\begin{align}
L_z = \frac{1}{3} + \frac{1}{2}\int_0^\pi d\theta \cos\theta\sin^2\theta x_\theta x^{-1}= (\xi_0^2-1)Q_1^0.
\end{align}

\subsection{All coefficients for $\b K^{eo}$, $m=0$}
\label{AppCoefm0}

For reference, we list below the expressions obtained for the coefficients needed to compute
$\b K^{eo}$ for $m=0$. Note that $Q_n^m\equiv Q_n^m(\xi_0)$.

\begin{align}
p_{11}&=-\frac{2i}{9 h^2}(s^2-1) \nonumber\\[0.2cm]
p_{11}^{(2)}&=i\frac{s^4-1}{225h^2}\frac{5\xi_0^2-4}{\xi_0^2}\nonumber\\[0.2cm]
p_{13}&=\frac{-4is^2(s^2-1)}{225\sqrt{14}h^2\xi_0^2} \nonumber\\[0.2cm]
p_{31}&= \frac{1}{s^2} p_{13}\nonumber\\[0.2cm]
u_{11}&= -i\left[1 + (s^2-1)L_z\right]\nonumber\\[0.2cm]
u_{11}^{(2)}&=-i\frac{s^2-1}{10}(\xi_0^2-1)\left[(2s^2+5)\frac{Q_2^0}{\xi_0}  - (s^2+10)Q_1^0 \right] \nonumber\\[0.2cm]
u_{13}&=\frac{-2is^2(s^2-1)}{25\sqrt{14}h^2} \left[ Q_1^0-\frac{3}{2}(5\xi_0^2-1)Q_3^0  \right] \nonumber\\[0.2cm]
u_{31}&=\frac{-3i\sqrt{14}}{4}(s^2-1)[ \xi_0^2 Q_3^0 - \xi_0 Q_2^0 - 2s^2(\xi_0^2-1)Q_3^0 ] \nonumber\\[0.2cm]
u_{33}&=-is^2\left[1+(s^2-1)(\xi_0^2-1)\frac{3}{2}(5\xi_0^2-1)Q_3^0 \right].\nonumber
\end{align}


\subsection{All coefficients for $\b K^{eo}$, $m=1$}
\label{AppCoefm1}

We also list below the expressions obtained for the coefficients needed to compute
$\b K^{eo}$ for $m=1$. For expressions involving Legendre functions of the second kind $Q_n^m(\xi_0)$ for $m=1$, we note that the proper branch cut in the complex plane must be taken. Correct expressions are given in App. \ref{AppLegendre}.

\begin{align}
p_{11}&=-\frac{2i}{9h^2}(s^2-1) \nonumber\\
p_{11}^{(2)}&=i\frac{s^4-1}{225h^2}\frac{5\xi_0^2-3}{\xi_0^2} \nonumber\\
p_{12}&=\frac{s^2(s^2-1)}{45\sqrt{15}h^2\xi_0^2} \nonumber\\
p_{13}&=\frac{-4is^2(s^2-1)}{225\sqrt{21}h^2\xi_0^2} \nonumber\\
u_{11}& =-i[ 1 + (s^2-1)L_x]\nonumber\\
u_{11}^{(2)}&=-i\frac{s^2-1}{20}\frac{\xi_0}{h^2}\left[ (s^2+10)\frac{\xi_0}{h}Q_1^1-(2s^2+5)Q_2^0  \right] \nonumber\\
u_{12}&=\frac{s^2(s^2-1)}{20\sqrt{15}h}\ ( Q_3^1 - Q_1^1)\nonumber\\
u_{21}&=-\frac{5(s^2-1)}{2\sqrt{15}}\frac{\xi_0}{h} Q_2^1 \nonumber\\
u_{22}&=-is^2 \nonumber\\
u_{13}&=-i\frac{s^2(s^2-1)}{25\sqrt{21}h} \left[ \frac{15\xi_0^2-11}{4}Q_3^1 - Q_1^1 \right] \nonumber\\
u_{31}&=-i\frac{\sqrt{21}}{24}(s^2-1)\frac{\xi_0}{h}[ 3(2s^2-1)\xi_0 Q_3^1 + 4Q_2^1 ] \nonumber\\
u_{32}&=-\frac{\sqrt{35}}{8}s^2(s^2-1)\frac{\xi_0^2}{h}Q_3^1 \nonumber\\
u_{33}&=-is^2\left[1-\frac{s^2-1}{8}\frac{\xi_0^2}{h}(15\xi_0^2-11)Q_3^1\right]\nonumber.
\end{align}

\subsection{Legendre Functions}
\label{AppLegendre}

All $Q_n^m$ of lowest order can be conveniently expressed in terms of $Q_0$ given by:
\begin{align}
Q_0^0=Q_0 = \frac{1}{2}\ln \frac{\xi+1}{\xi-1}=\text{acoth}(\xi)=\text{atanh}(e)
\end{align}
The others are
\begin{align}
Q^0_1 &= \xi Q_0 - 1 \\
Q^0_2 &= \frac{3\xi^2-1}{2} Q_0 -\frac{3\xi}{2}\\
Q^0_3 &= \frac{\xi (5\xi^2-3)}{2} Q_0  -\frac{15\xi^2-4}{6}\\[0.2cm]
Q^1_1 &= \frac{(\xi^2-1)Q_0-\xi}{\sqrt{\xi+1}\sqrt{\xi-1}}\\
Q^1_2 &= \frac{3\xi(\xi^2-1)Q_0-3\xi^2+2}{\sqrt{\xi+1}\sqrt{\xi-1}}\\
Q^1_3 &= \frac{3(\xi^2-1)(5\xi^2-1)Q_0 - \xi(15\xi^2 -13)}{2\sqrt{\xi+1}\sqrt{\xi-1}}\\[0.2cm]
Q^2_2 &= 3(\xi^2-1)Q_0
\end{align}
These are defined for $\xi$ on the complex plane minus the real interval between $-1$ and $1$. The factors $\sqrt{\xi+1}\sqrt{\xi-1}$ should be left separate (not expressed as $\sqrt{\xi^2-1}$) to give correct results for all $\xi$, in particular $\xi$ negative imaginary (for oblate spheroids). These functions should coincide with Maple and the type 3 Legendre in the Wolfram documentation.

\bibliography{Tmatrix}

\end{document}


\title{Supplementary Information for ``Approximate $T$-matrix and optical properties of spheroidal particles to third order in size parameter'' }

\author{Matt R. A. Majic}
\author{Luke Pratley}
\author{Dmitri Schebarchov}
\author{Walter R. C. Somerville}
\author{Baptiste Augui{\'e}}
\author{Eric C. Le Ru}\email{eric.leru@vuw.ac.nz}

\affiliation{The MacDiarmid Institute for Advanced Materials and Nanotechnology,
School of Chemical and Physical Sciences, Victoria University of Wellington,
PO Box 600 Wellington, New Zealand}

\date{\today}

%
\maketitle%

\onecolumngrid
 
Similar figures to those in the main text -- for the dipole matrix elements, other matrix elements, and cross sections -- are presented for different parameters. In the main text we used prolate spheroids of aspect ratio $h=3$ and relative refractive index $s=1.3$ or $s=1.3+0.2i$. Here we test prolate spheroids with $h=10$ and $s=1.3, 1.3+0.2i$, higher-index prolate spheroids with $h=3$, and $s=1.7$ or $s=\sqrt{-10+0.5i}$ (i.e. metallic, comparable to Silver at 530\,nm), and oblate spheroids with $h=1/3$ and $s=1.3, 1.3+0.2i$.

\section{High aspect ratio prolate spheroids}

\begin{figure}[h]
\includegraphics[width=18cm,clip=true,trim=0cm 0cm 0cm 0cm]{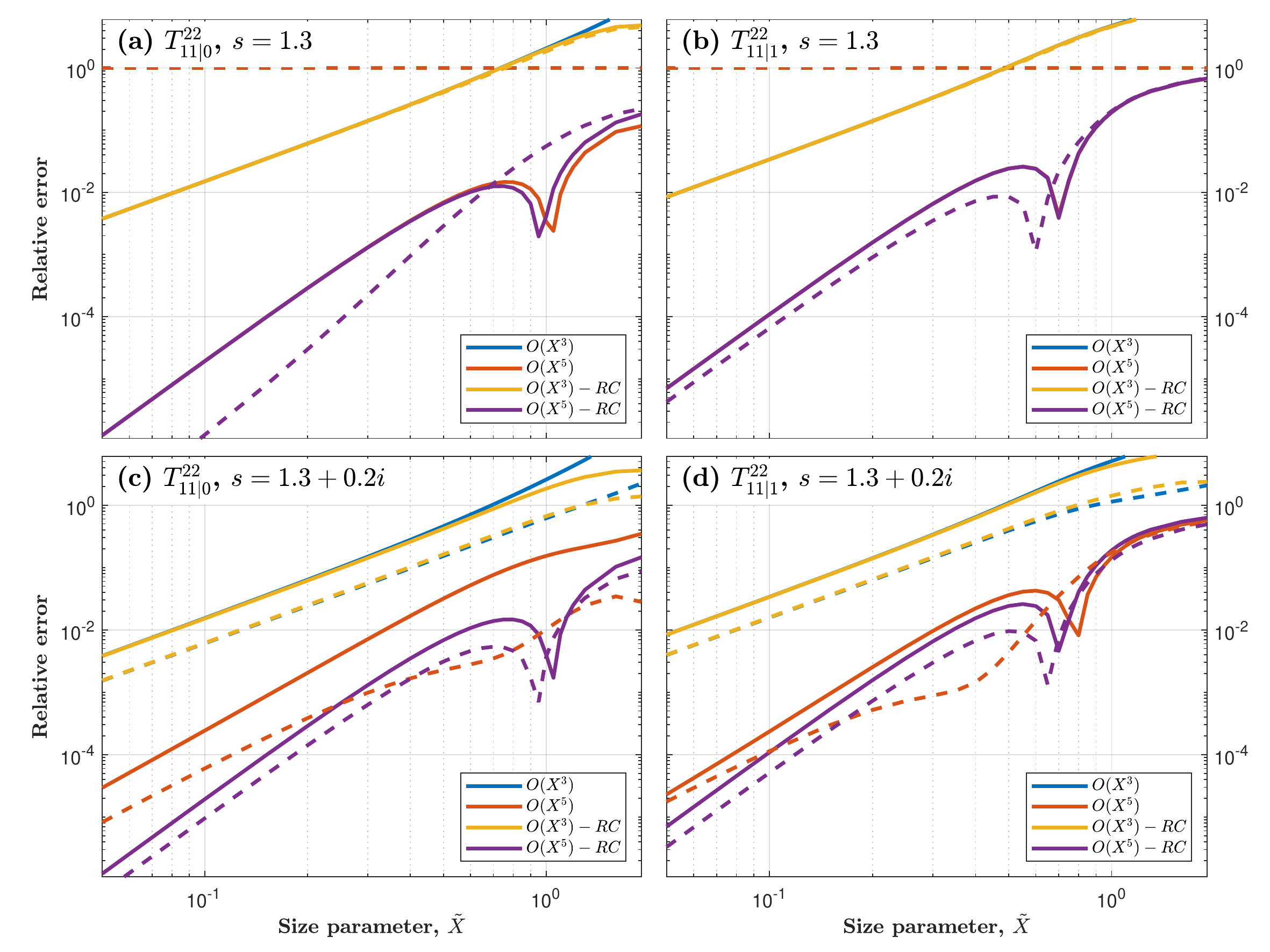}
\caption{Same as Figure 1 for prolate spheroids of aspect ratio $h=10$.}
\end{figure}

\begin{figure}
\includegraphics[width=18cm,clip=true,trim=0cm 0cm 0cm 0cm]{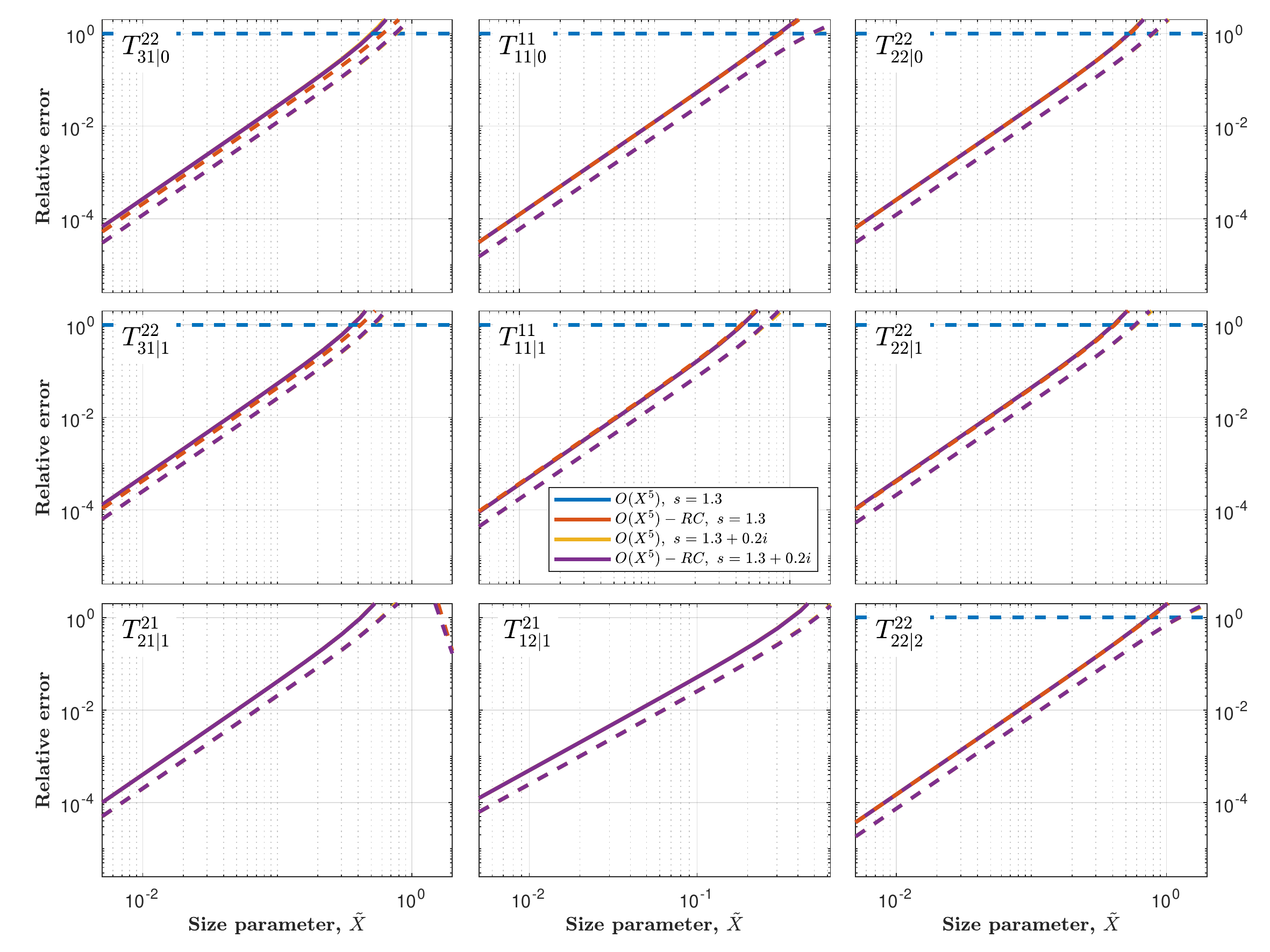}
\caption{Same as Figure 2 for prolate spheroids of aspect ratio $h=10$.}
\end{figure}

~\newpage
~\mbox{~}\\

\begin{figure}
\includegraphics[width=18cm,clip=true,trim=0cm 0cm 0cm 0cm]{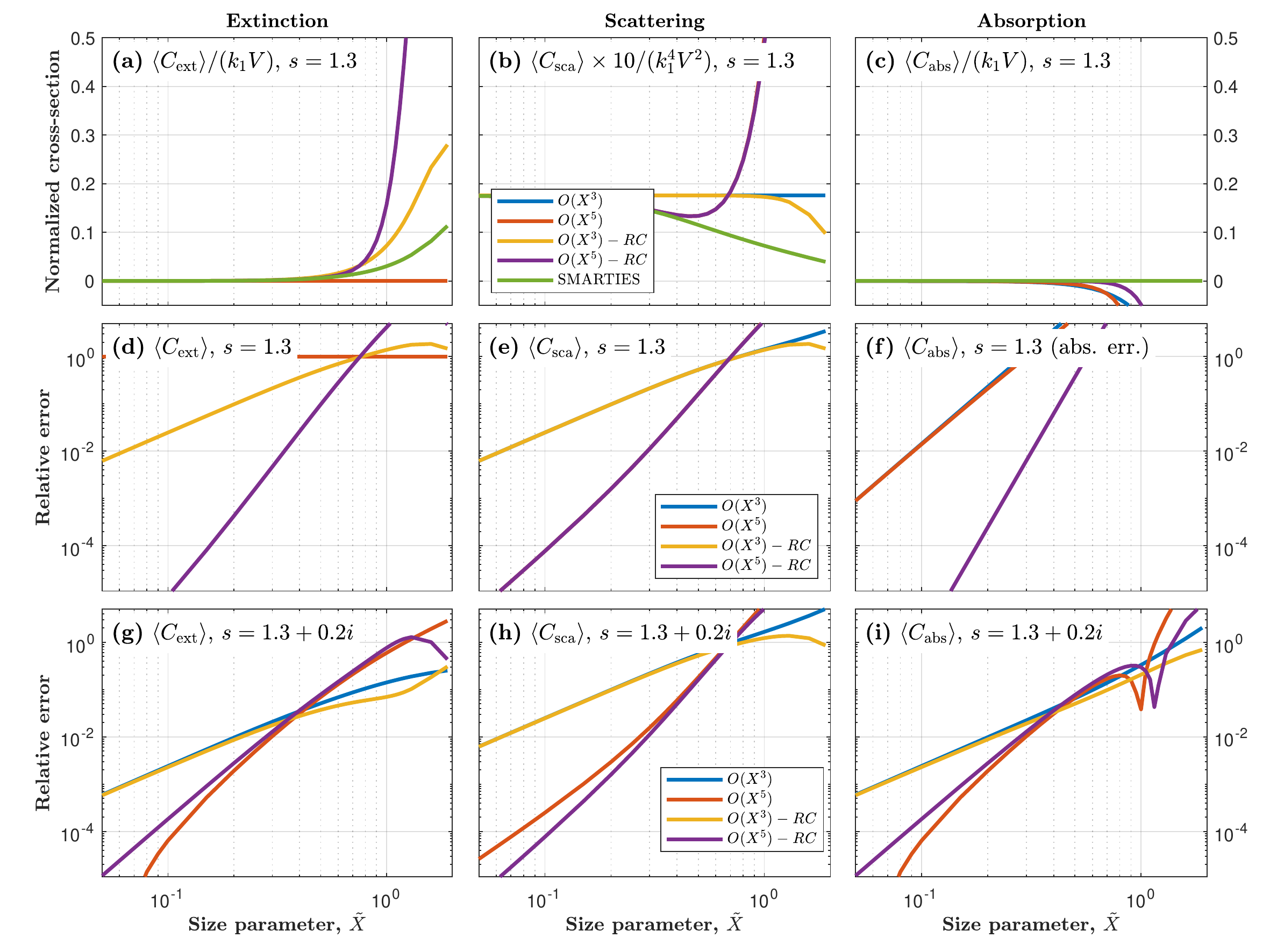}
\caption{Same as Figure 3 for prolate spheroids of aspect ratio $h=10$.}
\end{figure}

~\newpage
~\mbox{~}\\

\section{High-index prolate spheroids}
 
\begin{figure}[h]
\includegraphics[width=18cm,clip=true,trim=0cm 0cm 0cm 0cm]{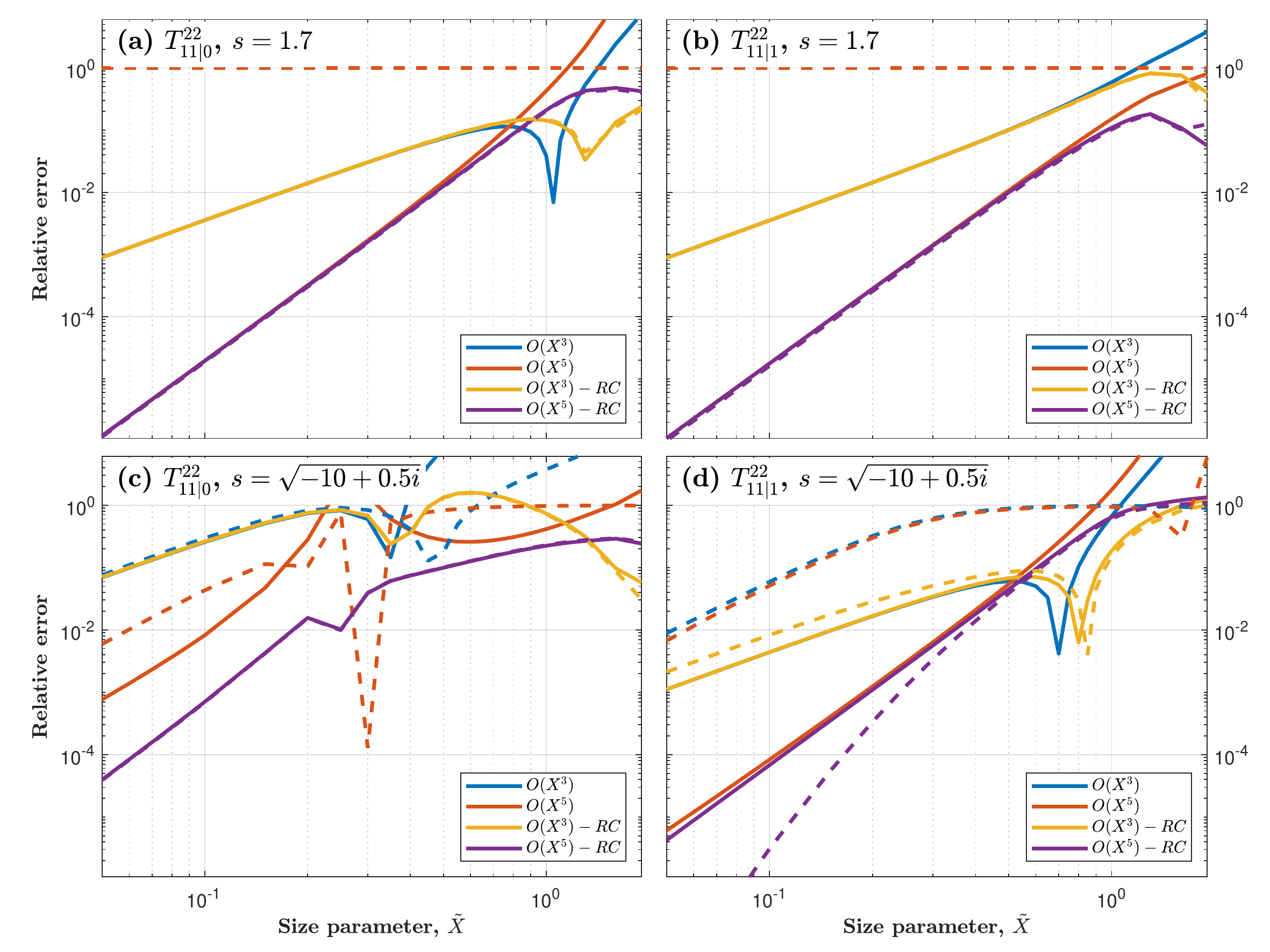}
\caption{Same as Figure 1 for different materials: prolate spheroids of aspect ratio $h=3$ either non-absorbing with $s=1.7$,
or metallic with $s=\sqrt{-10+0.5i}$.}
\end{figure}

~\newpage
~\mbox{~}\\

\begin{figure}
\includegraphics[width=18cm,clip=true,trim=0cm 0cm 0cm 0cm]{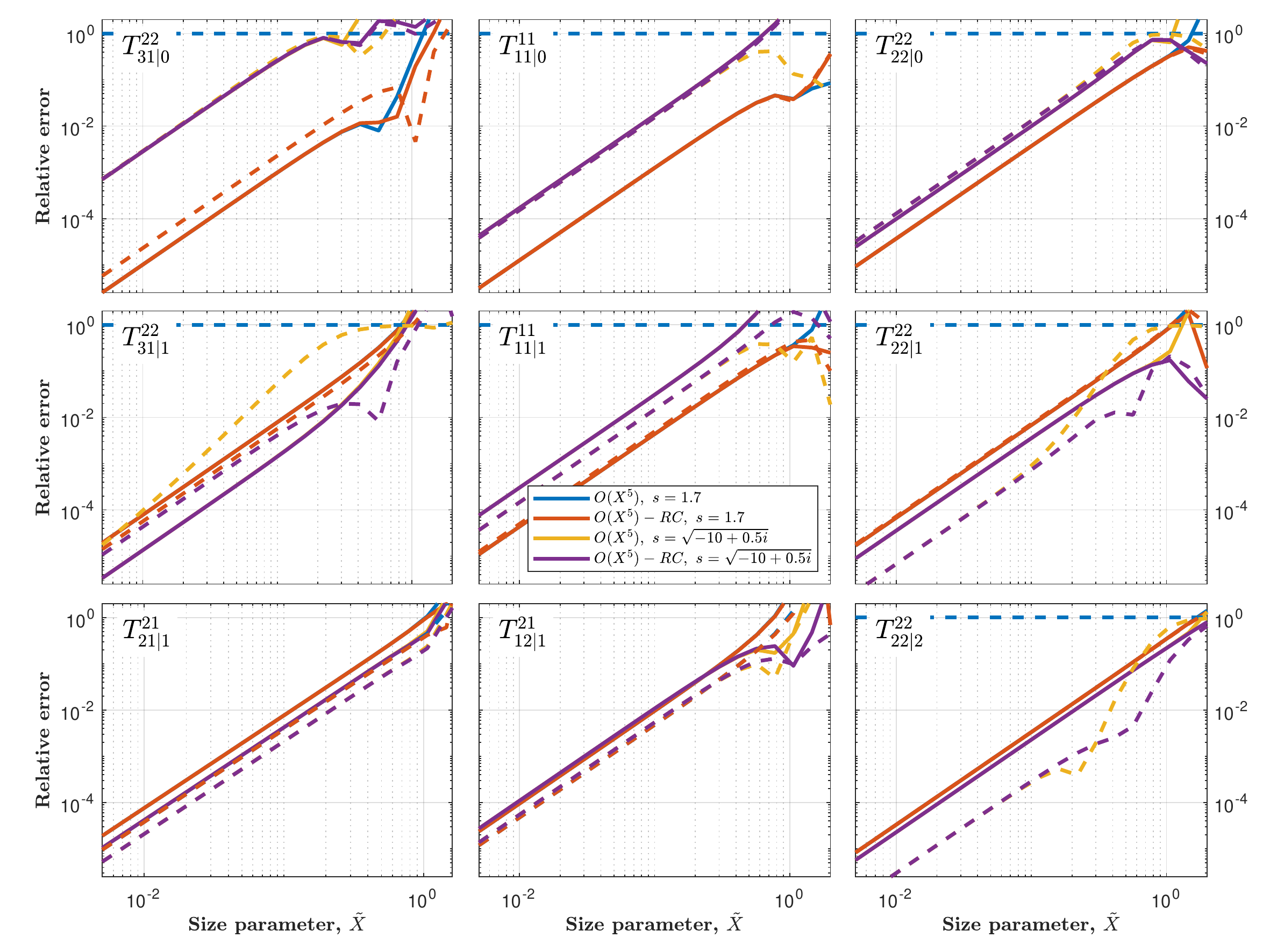}
\caption{Same as Figure 2 for different materials: prolate spheroids of aspect ratio $h=3$ either non-absorbing with $s=1.7$,
or metallic with $s=\sqrt{-10+0.5i}$.}
\end{figure}

~\newpage
~\mbox{~}\\

\begin{figure}
\includegraphics[width=18cm,clip=true,trim=0cm 0cm 0cm 0cm]{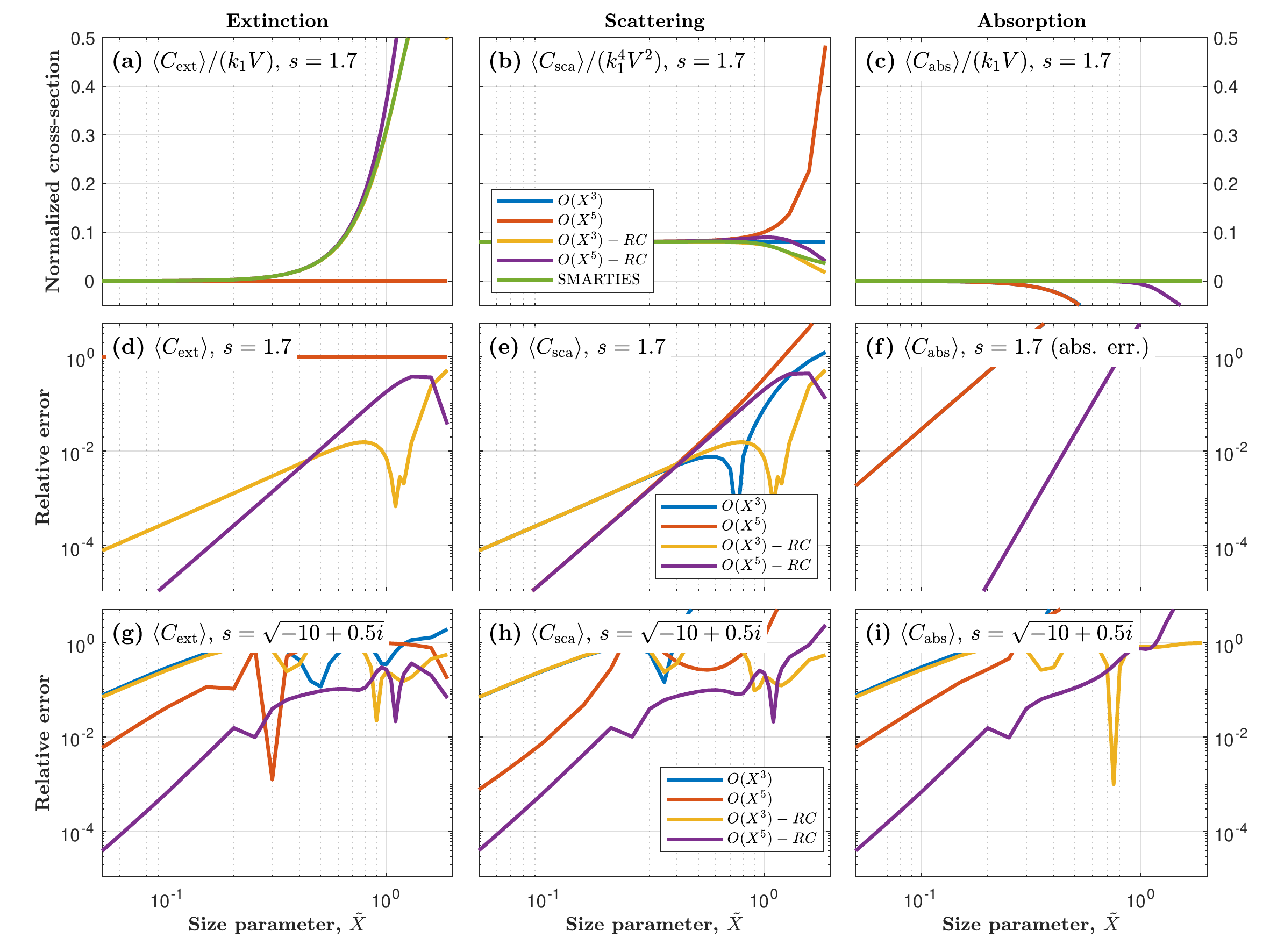}
\caption{Same as Figure 3 for different materials: prolate spheroids of aspect ratio $h=3$ either non-absorbing with $s=1.7$,
or metallic with $s=\sqrt{-10+0.5i}$.}
\end{figure}

~\newpage
~\mbox{~}\\

\section{Oblate spheroids}

\begin{figure}[h]
\includegraphics[width=18cm,clip=true,trim=0cm 0cm 0cm 0cm]{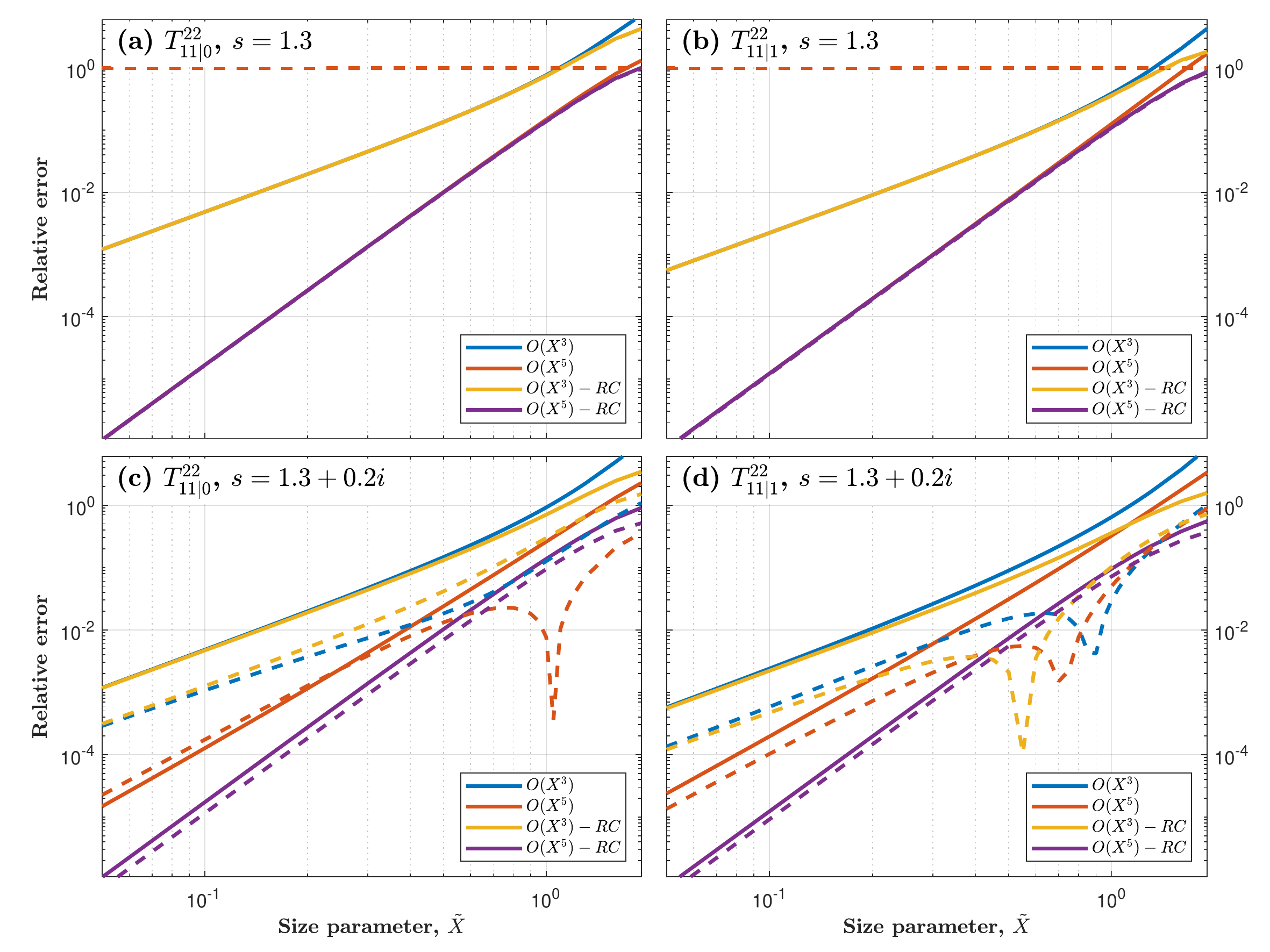}
\caption{Same as Figure 1 for oblate spheroids of aspect ratio $h=1/3$.}
\label{FigDipolarS1}
\end{figure}

~\newpage
~\mbox{~}\\

\begin{figure}
\includegraphics[width=18cm,clip=true,trim=0cm 0cm 0cm 0cm]{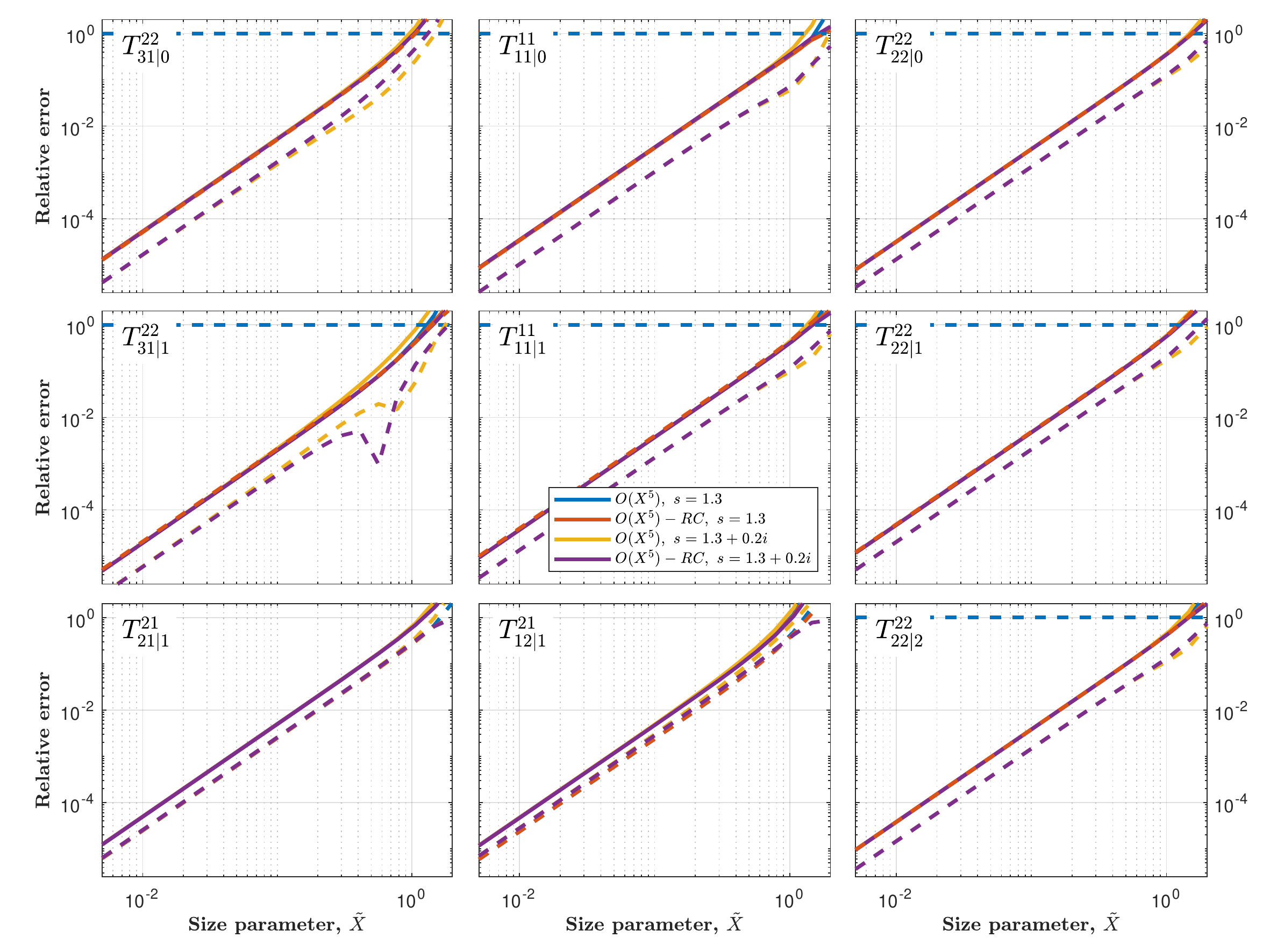}
\caption{Same as Figure 2 for oblate spheroids of aspect ratio $h=1/3$.}
\label{FigOthersS3}
\end{figure}

~\newpage
~\mbox{~}\\

\begin{figure}
\includegraphics[width=18cm,clip=true,trim=0cm 0cm 0cm 0cm]{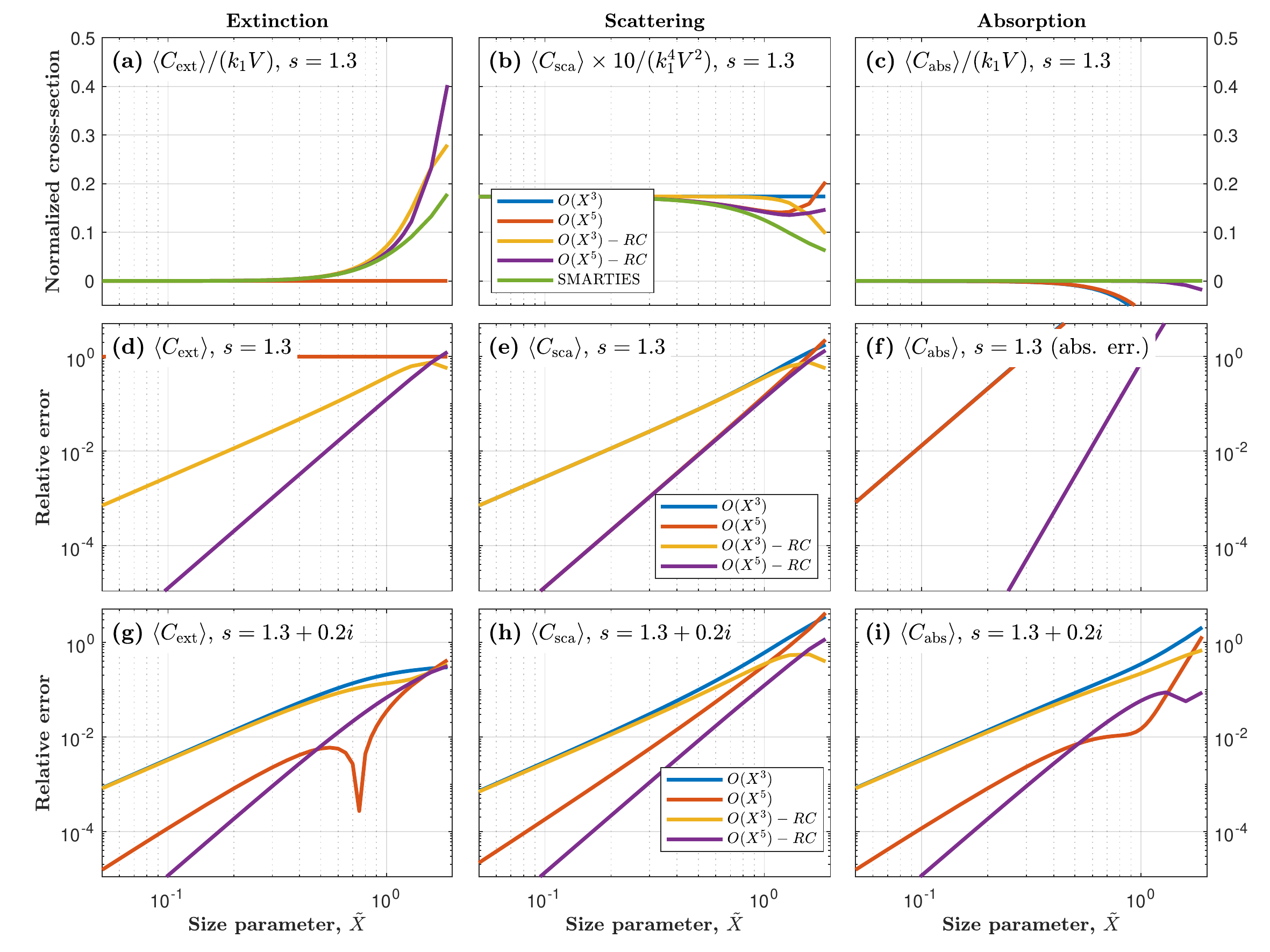}
\caption{Same as Figure 3 for oblate spheroids of aspect ratio $h=1/3$.}
\label{FigDerivedOblate}
\end{figure}
